\documentclass{aa}

\usepackage{graphics}

\newcommand{\solm}{$M_{\sun}$\ }
\newcommand{\solarb}{$L_{\sun}$\ }
\newcommand{\solar}{$L_{\sun}$\ }
\newcommand{\solars}{$L_{\sun}$\ }

\newcommand{\apjj}[2]{ApJ #1, #2}
\newcommand{\apjs}[2]{ApJS #1, #2}
\newcommand{\apjjl}[2]{ApJ #1, L#2}
\newcommand{\ajj}[2]{AJ #1, #2}
\newcommand{\asa}[2]{A\&A #1, #2}
\newcommand{\asas}[2]{A\&AS #1, #2}
\newcommand{\anrev}[2]{ARA\&A #1, #2}

\newcommand{\mn}[2]{MNRAS #1, #2}

\begin{document}

\thesaurus{11(11.03.1; 11.03.4 Abell~262; 11.05.2; 11.09.2; 11.09.4)}

\title{
Near-infrared adaptive optics observations of galaxy clusters:\\
\object{Abell~262} at z=0.0157,
\object{J1836.3CR} at z=0.414,
and
\object{PKS~0743-006} at z=0.994}

\author{
Hackenberg, W.\inst{1,2},
Eckart, A.\inst{1,3}, 
Davies, R.I.\inst{1}, Rabien, S.\inst{1}, Ott, T.\inst{1},
Kasper, M.\inst{4}, Hippler, S.\inst{4}, 
\and 
A. Quirrenbach\inst{5}}

\institute{Max-Planck-Institut f\"ur extraterrestrische Physik (MPE),
D-85740 Garching, Germany
\and European Southern Observatory (ESO),
 D-85748 Garching, Germany
\and Universit\"at zu K\"oln, D-50937 K\"oln, Germany
\and Max-Planck-Institut f\"ur Astronomie (MPIA), D-69117 Heidelberg, Germany
\and University of California, San Diego, La Jolla, CA 92093-0424, USA}

\date{Received / Accepted 15. May 2000}

\titlerunning{Near-infrared adaptive optics observations of galaxy
clusters}
\authorrunning{W. Hackenberg et al.}

\maketitle

\begin{abstract}

We report on high angular resolution near-infrared (NIR) observations
of three galaxy clusters at different redshifts using adaptive optics (AO).
In the case of the barred spiral \object{UGC~1347} in Abell~262 we presented
the first AO results obtained using a laser guide star.
The observations have been carried out with the MPE/MPIA adaptive
optics laser guide star system ALFA and the ESO AO system ADONIS
combined with the SHARP~II+ camera built at MPE.
The three clusters are well suited for high resolution investigations since
bright field stars for tip-tilt or wavefront sensing are located close to
the line of sight to cluster galaxies.
In summary our high angular resolution NIR data combined with other information
clearly indicates star formation activity or interaction between
cluster members at all three redshifts.
The results and implications for future high angular
resolution adaptive optics observations
are discussed in the framework of current galaxy and
cluster evolution models.

For two barred galaxies in the Abell~262 cluster, \object{UGC~1344} and UGC~1347,
we interpret our NIR imaging results in combination with published
radio, far-infrared, and H$\alpha$ data in the framework of a star formation
model.
In addition to the
star-forming resolved NIR nucleus in UGC~1347 we found
a bright and compact region of recent and enhanced star formation at one tip of the bar.
The $L_{\rm K}/L_{\rm Lyc}$ ratio as well as the V~$-$~K color of that region
imply a starburst that happened
about 10$^7$ years ago.
For UGC~1344 we find that the overall star formation activity is low
and that the system is deficient in fuel for star formation.

The importance of star formation in galaxy clusters
is also supported by a comparison of seeing corrected
nuclear bulge sizes of a sample of
spiral galaxies within and outside the central HI deficient
zone of the Abell~262 and \object{Abell~1367} clusters.
We find that the galaxies inside the Abell radii of both clusters
show a tendency for more compact bulges than those outside.
This phenomenon could be due to increased star formation activity
triggered by interactions of cluster members inside the Abell radius.

The star formation activity in the two higher redshift clusters
J1836.3CR and PKS~0743-006 is investigated via comparison to
GISSEL stellar population models in JHK two-color-diagrams.
While J1836.3CR is consistent with an evolved cluster, the objects in the
field of PKS~0743-006 show indications of more recent star formation
activity.
The central object in J1836.3CR shows a radial intensity profile
that is indicative for cD galaxies in a rich cluster environment.
Extended wings in its light distribution may be consistent with
recent or ongoing galaxy-galaxy interaction in this cluster.


\end{abstract}

\section{Introduction}

Adaptive optics systems using natural and laser guide stars are an
important observational tool that allow large ground based
telescopes to operate at or close to the diffraction limit.
Considerable improvements and successes have been obtained in
installing such systems at several sites
(Davies et al. \cite{Davies1999},
Davies et al. \cite{Davies1998},
Glindemann et al. \cite{Glindemann},
Quirrenbach et al. \cite{Quirrenbach},
Drummond et al. \cite{Drummond},
Hubin \cite{Hubin},
Max et al. \cite{Max},
Arsenault et al. \cite{Arsenault}).
However, it remains
challenging to use them efficiently especially in the field of
extragalactic observations.
We have concentrated on three galaxy clusters at different redshifts
for which adaptive optics observations were possible due to the presence
of sufficiently bright reference stars in the corresponding fields.

The galaxy cluster Abell~262
(R.A.(2000) = 01$^h$52.1$^m$, DEC(2000) = 35$^\circ$40$'$)
is one of the most conspicuous condensations
in the Pisces-Perseus super cluster. It has a systemic velocity of 4704
km~s$^{-1}$ ($z=0.0157$, Giovanelli and Haynes \cite{Giovanelli1985})
and an Abell radius of $r_A$ = 1.75$^\circ$.
It has been extensively studied in
X-rays and in the radio. It is a spiral-rich cluster, characterized by
the presence of a central X-ray source positioned on the D galaxy
NGC~708 right at the center of the cluster.
The distribution of galaxies in projection on the sky as well as in
redshift space
have been studied by Melnick and Sargent (\cite{Melnick}),
Moss and Dickens (\cite{Moss1977}),
Gregory et al. (\cite{Gregory}), and Fanti
et al. (\cite{Fanti}).
The large number of spirals in this cluster as well as the presence of
a central X-ray source and its low redshift make Abell~262 an ideal
candidate to study the properties of member galaxies, such as HI
content and star formation activity.
As in many other rich galaxy clusters the member galaxies of Abell~262 show
an HI deficiency towards the center of the cluster. For Abell~262 this
phenomenon
has been investigated by Giovanelli et al. (\cite{Giovanelli1982}),
Giovanelli \& Haynes (\cite{Giovanelli1985}) and others.

We used the new MPIA-MPE ALFA adaptive optics
system at the Calar Alto 3.5~m telescope
to observe two of the Abell~262 cluster members --~UGC~1344 and UGC~1347~--
at subarcsecond resolution.
For UGC~1347 the observations were carried out using the ALFA laser guide star
(LGS) and a nearby natural guide star (NGS) for tip-tilt correction. To our
knowledge UGC~1347 is the first extragalactic source for which LGS assisted
observations have been performed.
For the UGC~1344 observations we used a nearby NGS as a wavefront reference.

As the two higher redshift clusters we selected J1836.3CR
(R.A.(2000) = 13$^h$45$^m$, DEC(2000) = $-$00$^\circ$53$'$)
at a redshift of $z=0.414$ (Couch et al. \cite{Couch})
and an area around the quasar PKS~0743-006 at a redshift of
$z=0.994$ (Hewitt \& Burbidge \cite{Hewitt}).
Both fields contain bright guide stars with $m_{\rm V} \sim 12$~mag.
that can be used as natural guide stars for adaptive optics observations.

Couch et al. (\cite{Couch}) presented a catalogue of faint southern galaxy
clusters identified on high-contrast film derivatives of a set of
Anglo-Australian Telescope photographic plates. The cluster J1836.3CR is one of them.
A bright star ($m_{\rm V}$ = 12~mag) is located about 60$''$ north of 4 prominent
cluster members for which redshifts have been determined.
For three of four galaxies spectroscopy (Couch et al. \cite{Couch})
indicates a redshift of $z=0.415 \pm 0.003$
 and one galaxy has a redshift of $z=0.319$.
Couch et al. (\cite{Couch}) use a cluster redshift of $z=0.414$.
At this redshift 1$''$ corresponds to a linear distance of  6.8~kpc.

PKS~0743-006 is a quasar
(R.A.(2000) = 07$^h$45$^m$53.37$^s$, DEC(2000) = $-$00$^\circ$44$'$11.4$''$)
of visual magnitude $m_{\rm V}$ = 17.1~mag
at a redshift $z=0.994$ (Hewitt \& Burbidge \cite{Hewitt}).
The radio spectrum has a convex shape with possible variability
around the peak occurring between 5 and 10 GHz.
Tornikoski et al. (\cite{Tornikoski}) find this source strongly variable at 90 GHz.
Variability by a few tenths of a magnitudes is also reported in the NIR
(White et al. \cite{White}).
On the milliarcsecond angular resolution scale at cm wavelengths
this object shows a classical core-jet structure
(Stanghellini et al. \cite{Stanghellini}).
Within the errors, the whole radio flux density is
accounted for by this structure.
A natural guide star for adaptive optics observations is located at
only about 12.2$''$ northeast of the quasar.

In section 2 we describe the observations and data reduction as well
as the adaptive optics systems we used. In section 3 we present
the observational results and
the data analysis for UGC~1347 (section 3.1) and for UGC~1344 (section 3.2)
 in conjunction with data available in the literature.
In section 3.3 we outline the results we obtained for a sample
of 11 spiral galaxies in the Abell~262 cluster and 15 spiral galaxies
in the Abell~1367 cluster.
In section 4 then we discuss the star formation activity in the observed
cluster galaxies and give a summary and conclusions in section 5.

\section{Observations and data reduction}

Our new  high spatial resolution observations  in Abell~262 were
carried out using the OMEGA-CASS  camera
mounted to the laser guide star adaptive optics system ALFA at the Calar
Alto 3.5~m telescope.
The observations for the two higher redshift clusters were obtained with the
ESO AO system ADONIS. In the following we give a brief description of the two
systems.

\subsection{ALFA}

The performance goal of ALFA is to achieve a 50\% Strehl-ratio at 2.2\,$\mu$m under
average seeing conditions (0.9$''$), with good sky coverage.
The adaptive optics (Glindemann et al. \cite{Glindemann}) and
the sodium laser guide star (Quirrenbach et al. \cite{Quirrenbach}, Davies et al. \cite{Davies1998})
have been designed and built as a joint project
between MPIA in Heidelberg and MPE in Garching, both in Germany.
The system is installed at the German/Spanish 3.5 m telescope on Calar
Alto near Almeria, Spain.

The laser used for generating the artificial guide star is a high
power continuous-wave dye laser. It is 
installed in the coud\'e lab of the telescope, and the laser beam is fed
along the coud\'e train until it is picked off near the primary mirror
and directed into a 50-cm launch telescope. The launched laser power is around
3\,W, and produces a $m_{\rm V} =$~9--10~mag sodium guide star.
The tip-tilt correction is achieved using a natural guide
star, currently with a limiting magnitude $m_{\rm V} \sim 15$~mag.
The laser can be used for high order wavefront correction.
In the wavefront sensor there are several lenslet arrays which can be interchanged, and the
positions of the resulting laser beacon centroids in the
Shack-Hartmann sensor are determined and used to derive coefficients of
Zernike or Karhunen-Loeve modes which are then used to control a 97-actuator
deformable mirror.
The loop was closed on the laser guide star in September 1997, and it was
first used to improve an image in December 1997 at a sampling rate of 60~Hz
and correcting 7 modes plus tip and tilt.

OMEGA-CASS is a near-infrared camera for the Cassegrain focus of the
3.5 m telescope at Calar Alto, which is specialized for use at high
spatial resolution  and has been developed at the MPIA, Heidelberg.
It is based around a Rockwell 1024$^2$ pixel HAWAII array, and has
capabilities for broad and narrow band imaging, spectroscopy, and
polarimetry over the 1.0--2.5$\,\mu$m wavelength range.
When used in conjunction with ALFA ($f$/25), the pixel scales available
are 0.04$''$, 0.08$''$, and 0.12$''$ per pixel.

\subsection{ADONIS}

For the observations of the two higher redshift clusters we used the
ESO adaptive optics system ADONIS (Beuzit et al. \cite{Beuzit}). This system is operated at ESO's
3.6 m telescope at La Silla, Chile, and includes the SHARP~II+
camera built at MPE.
The atmospheric wavefront distortions are measured with a
Shack-Hartmann sensor at visible wavelengths and are corrected by
a deformable mirror with 52 piezo actuators. This mirror is driven by a
closed control loop with a correction bandwidth of up to
17~Hz. The natural guide star within the near-infrared isoplanatic
patch must be brighter than about m$_{\rm V}=13$~mag.
The SHARP II+ camera (Hofmann et al. \cite{Hofmann}, Eisenhauer et al. \cite{Eisenhauer}) is
based on a 256$^2$ pixel NICMOS III detector. The
wavelength range of our observations covers the atmospheric  J, H and K bands.
Compared with the Johnson K band, we used a somewhat narrower K$'$ filter
(1.99 -- 2.32 $\mu$m)  in order
to reduce the thermal background.

\subsection{The data}
Goal of our investigation was to exploit structural information 
on galaxy cluster members from adaptive optics and seeing limited images
and interpret the results making use of all available quantities and known
correlations.
The photometric quality of the data is of the order of 0.10$^m$ in the K- 
and H-band and 0.15$^m$ in the J-band.
The sources were mainly selected on the basis of availability during 
the allocated observing time, the presence of bright AO reference stars, and
the availability of additional literature data. The sample described at the
end of section 2.3.1 was selected on the basis HI deficiency and beeing
located within or outside the cluster`s Abell radii.

\subsubsection{Abell~262 and Abell~1367}

The broad-band J, H, K images as well as first K-band adaptive optics data
of the cluster member UGC~1347
were taken on November 10 and 12, 1997.
K-band AO data of UGC~1344 and UGC~1347
as well as the direct imaging data of other
cluster members were obtained on December 6 and 7, 1997.
Many individual 5-second  exposures were taken in all three bands and coadded
after sky subtraction, flat-fielding and correcting for bad pixels.
The integration times, pixel scales and angular resolutions
of the final co-added images are listed in Tab.~\ref{ww01}.

\small
\begin{table*}
\caption{\label{ww01} Observing parameters for Abell 262 and Abell 1367.}
\begin{center}
\begin{tabular}{cccccccccc}\hline \hline
date     &source&band&scale    &t$_{\rm int}$&$N_{\rm set}$&$N_{\rm images}$&$t_{\rm total}$&resolution&mode\\
in 1997  &      &    &($''$/pixel)& (s)   &         &            & (s)        & FWHM  ($''$)&    \\ \hline
Nov. 10-12  &UGC~1347  &  J  & 0.12 & 5        & 10 &15 & 750 & 1.1       & direct  \\
            &UGC~1347  &  H  & 0.12 & 5        &  8 &15 & 600 & 1.0       & direct  \\
            &UGC~1347  &  K  & 0.12 & 5        &  8 &15 & 600 & 0.9       & direct  \\
Dec. 6-7    &UGC~1347  &  K  & 0.12 & 5        &  2 &15 & 150 & 1.2       & open loop  \\
            &UGC~1347  &  K  & 0.12 & 5        &  5 &15 & 375 & 0.4       & closed loop  \\
            &UGC~1344  &  K  & 0.08 & 5        &  2 &15 & 150 & 0.9       & open loop  \\
            &UGC~1344  &  K  & 0.08 & 5        &  5 &15 & 375 & 0.4       & closed loop  \\
            &all others&  K  & 0.12 & 4 - 5    &  2 &15 & 150 & 1.2 - 1.4 & direct  \\
\hline 
\end{tabular}
\end{center}
\end{table*}
\normalsize
\small
\begin{table*}
\caption{\label{ww02} Observing parameters for J1836.3CR.All on-source
integrations were carried out with the AO-loop closed.}
\begin{center}
\begin{tabular}{cccccccc}\hline \hline
date  & band & scale  & $t_{\rm int}$ & $N_{\rm images}$ & $N_{\rm set}$ & $t_{\rm total}$ & resolution \\
in 1996 &      &($''$/Pixel) & (s)       &       &       & (s)         &
FWHM ($''$) \\
\hline
Apr. 26   & K$'$    & 0.10  & 60    & 5     & $4\times2$    & 2400  & 0.20\\
        & H     & 0.10  & 60    & 5     & 4             & 1200  & 0.20\\
        & J     & 0.10  & 60    & 5     & 4             & 1200  & 0.30\\
Apr. 27   & K$'$    & 0.05  & 60    & 5     & $1\times5$    & 1500  & 0.18\\
        & H     & 0.05  & 60    & 5     & $1\times4$    & 1200  & 0.20\\
Apr. 29   & K$'$    & 0.05  & 60    & 5     & 2             & 600   & 0.18\\
        & J     & 0.05  & 60    & 5     & 4             & 1200  & 0.25\\
May 5    & K$'$    & 0.05  & 60    & 5     & 3             & 900   & 0.18\\
        & H     & 0.05  & 60    & 5     & 3             & 900   & 0.20\\
        & J     & 0.05  & 60    & 5     & 2             & 600   & 0.25\\
\hline
\end{tabular}
\end{center}
\end{table*}
\normalsize
\small
\begin{table*}
\caption{\label{ww03} Observing parameters for PKS0743-006. All on-source
integrations were carried out with the AO-loop closed.}
\begin{center}
\begin{tabular}{cccccccc}\hline \hline
date & band & scale & $t_{\rm int}$ & $N_{\rm images}$ & $N_{\rm set}$ & $t_{\rm total}$ & resolution\\
in 1996 & &($''$/Pixel) & (s) & & & (s) & FWHM ($''$) \\
\hline
Apr. 26   & K$'$    & 0.10  & 30-60 & 5     & $4\times3$    & 3000  & 0.20\\
        & H     & 0.10  & 20    & 10    & 4             & 800   & 0.20\\
        & J     & 0.10  & 30    & 5     & 4             & 600   & 0.30\\
Apr. 27   & K$'$    & 0.10  & 60    & 3     & 3             & 540   & 0.20\\
        & K$'$    & 0.05  & 60    & 5     & 2             & 600   & 0.20\\
Dec. 20  & K$'$    & 0.10  & 60    & 5     & 4             & 1200  & 0.20\\
        & H     & 0.10  & 60    & 5     & 2             & 600   & 0.25\\
        & J     & 0.10  & 60    & 5     & 1             & 300   & 0.40\\
Dec. 21  & K$'$    & 0.05  & 60    & 5     & 8             & 2400  & 0.20\\
        & H     & 0.05  & 60    & 5     & 1             & 300   & 0.25\\
\hline
\end{tabular}
\end{center}
\end{table*}
\normalsize
Calibration of the NIR data was accomplished by observation
of the standard star $\xi$$^2$Ceti.
Sky data were taken separately 120$''$ east of UGC~1344 and UGC~1347.
For the other cluster members a median sky was obtained from the 4 to 5 settings
taken with different offset positions from the target sources.
The galaxies UGC~1344 and UGC~1347 had sufficiently bright reference stars
nearby to observe them with the ALFA adaptive optics system.
In order to estimate the image improvement we took data in open loop before and
after the closed loop exposures.
For UGC~1347
the wavefront data on the laser guide star
were taken at a sampling rate of 60~Hz through a 3$\times$3 lens\-let array
with field sizes of 3$''$ diameter.
Correcting for a total of 7 Zernike modes plus tip and tilt
a disturbance rejection bandwidth of up to 5~Hz was achieved.
The tip and tilt information was derived from a nearby natural reference star.

\begin{figure}
\resizebox{\hsize}{!}{\includegraphics{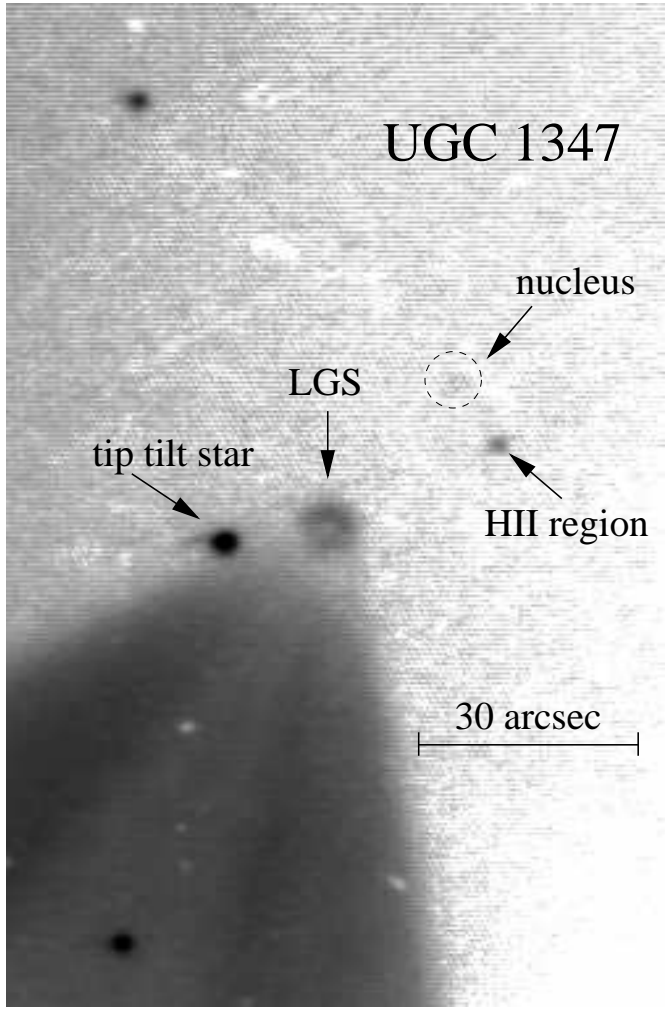}}
\caption{ALFA TV-guider image of UGC~1347 plus laser. The telescope was focused
at infinity, i.e., on natural stars, resulting in a defocused image of
the sodium LGS.}
\label{fig01}
\end{figure}

In Fig.~\ref{fig01} we show an image through the TV-guider shortly after
the LGS-supported AO observations of UGC~1347 were made. 
The image is
focused on the stars -- so the LGS appears as a defocused image at the tip of the
Rayleigh cone. We had placed the LGS between the tip-tilt reference star
and the nucleus of UGC~1347, such that the image of the star which
is also in the field of view of the NIR camera could be used as the
point spread function (PSF) with the same degree of correction as the nucleus
and most of the galaxy.
On the reference star for UGC~1344 the loop was closed using the same lens\-let
array and a camera frame rate of 100 to 200~Hz
resulting in a slightly higher rejection bandwidth.
Although we did not reach the diffraction limit due to the low
sampling rate and the small number of subapertures, definite
improvements in angular resolution were achieved. The corresponding
full-width-half-maximum (FWHM) values are given in Tab.~\ref{ww01}.
In the case of UGC~1347 the image improvement could independently be
monitored
via a star in the same field at approximately the same separation from
the reference star as the target object.
The two stellar images agreed very well with each other indicating that
all the sources were well within the isoplanatic patch and that the
images of the reference stars can safely be taken as the PSF
to clean the galaxy images.

In addition to the adaptive optics data we took seeing limited images with
exposure times of 10 minutes each of 9 galaxies in Abell~262 and 15 spirals
in Abell~1367 (see section 3.3 for further details).

\subsubsection{J1836.3CR and PKS~0743-006}

We observed these clusters using the SHARPII+ camera together with the
ESO adaptive optics system ADONIS on the 3.6 m telescope on La Silla, Chile.
The observations were conducted during the nights from April 26 till May 1, 1996.
A bright star was used to lock the AO system.
We took a series of 60 second exposures in a dither mode
in the near-infrared J, H, and K$'$ bands,
using pixel scales of 0.05 and 0.10~"/pixel.
The total integration
times and the angular resolution measured on a PSF reference are
listed in  Tab.~\ref{ww02}  and Tab.~\ref{ww03}.

\section{Results}

In the following we will present the results obtained for the two
galaxies UGC~1347 and UGC~1344 that were observed with the ALFA
adaptive optics system as well as for a sample of galaxies located
in the inner and outer part of the Abell~262 and Abell~1367 clusters.
We also describe source properties at other wavelengths
as well as quantities that we derived from them.
The description of this derivation is given in detail for UGC~1347.
For UGC~1344 we have used the same approach and only summarize the results.
We regard the corresponding analysis as an important consistency check between 
our own data and the data and correlations available in the literature.
Data on external galaxies even at lower or medium redshift will always be
sparse and it is required to make use of all the knowledge available to
allow for a full comparison to what is known in the local universe and at
different redshifts.

Although our NIR data has a subarcsecond resolution we extracted
K-band and H$\alpha$ fluxes
in larger apertures to conduct a starburst analysis
in section 4. The reason for this is that the radio and bolometric
luminosities for individual source components especially the
nucleus, disk, and  southern component in UGC~1347  can only be estimated
indirectly and can probably only be attributed to larger regions.
We have chosen a circular aperture of 7.2$''$ diameter corresponding to
a linear size of 2.2~kpc.

\subsection{UGC~1347}

UGC~1347 is an almost face-on SBc galaxy located at
R.A.(2000) = 01$^h$52$^m$45.9$^s$ and DEC(2000) = 36$^\circ$37$'$09$''$ approximately
57$'$ north of the center of Abell~262, well within the region in which the
largest amount of HI deficiency is observed.
There is a bright field star (PPM~1111, $m_{\rm V}$ = 11.5~mag) located about 37$''$
to the southeast of the galaxy. The HI content
of UGC~1347 was first studied by Wilkerson (\cite{Wilkerson}).
Velocity fields and intensity maps were obtained in HI by
Bravo-Alfaro (\cite{Bravo-Alfaro})
and
in  H$\alpha$ by Amram et al. (\cite{Amram}).
Amram et al. (\cite{Amram}) quote an inclination of  $i=30^\circ$.
Oly and Israel (\cite{Oly}) measured the 327~MHz radio continuum
flux density of UGC~1347, and the far-infrared flux densities as
measured by IRAS can be found in the IRAS point source catalogue
(Lonsdale et al. \cite{Lonsdale}).
The HI and H$\alpha$ data indicate a systemic velocity of UGC~1347
of 5524~km~s$^{-1}$ (Wilkerson \cite{Wilkerson})  and 5478~km~s$^{-1}$ (Amram et al. \cite{Amram}),
respectively. Here we assume that the difference of approximately
800~km~s$^{-1}$ between the cluster
velocity  of 4704~km~s$^{-1}$ and the systemic velocity is due to the
motion of the galaxy within the cluster. We therefore adopt
for UGC~1347 the cluster distance of 63~Mpc or a redshift of $z=0.0157$
assuming $H_0$ = 75~km~s$^{-1}$~Mpc$^{-1}$.
At this distance 1$''$ corresponds to about 310~pc.

\subsubsection{Near-infrared emission from UGC~1347}

\begin{figure}
\resizebox{\hsize}{!}{\includegraphics{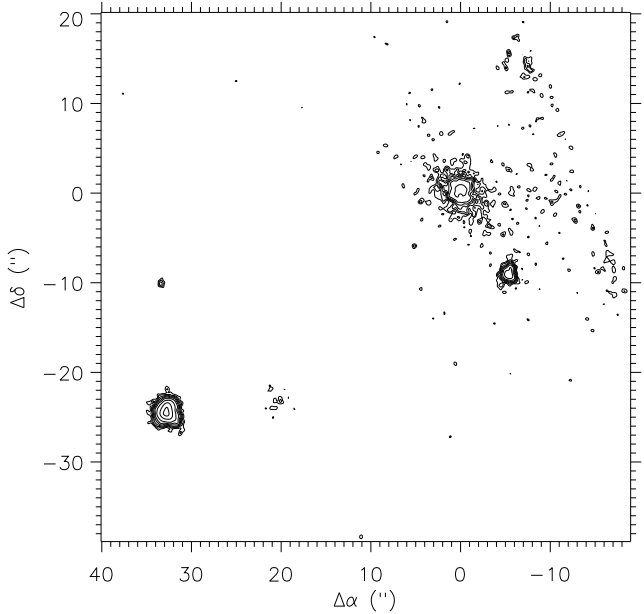}}
\caption{
AO-corrected K-band image of UGC~1347 with a resolution of 0.4$''$.
The brightest object to the southeast is the tip-tilt guide star.
The lowest contour level corresponds to 16.0~mag/arcsec$^2$.}
\label{fig02}
\end{figure}

\begin{figure}
\resizebox{\hsize}{!}{\includegraphics{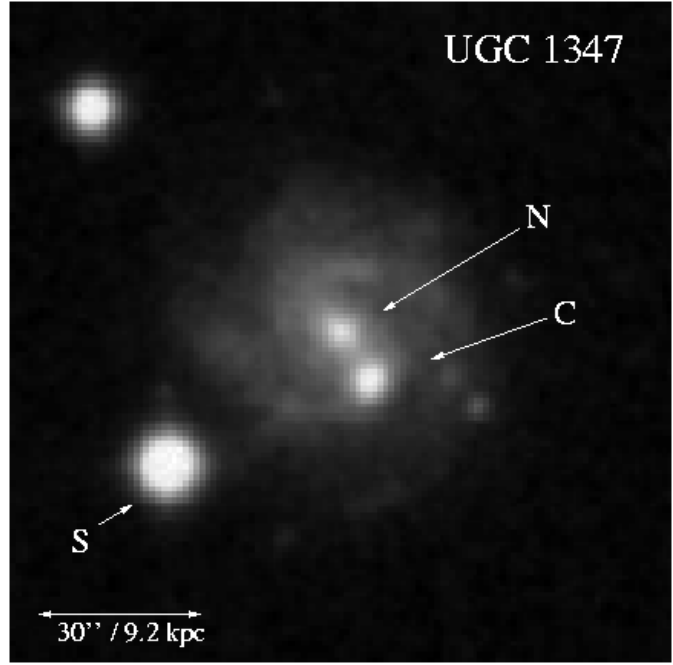}}
\caption{V-band image of UGC~1347 from the Palomar digitized sky survey.
The nucleus (N), the compact, bright off-nuclear source (C),
and the tip-tilt guide star (S) are labeled.}
\label{fig03}
\end{figure}

The NIR emission of UGC~1347 is dominated by two almost
equally bright components at a separation of
8.85$''$ (or 2.74~kpc) oriented approximately north-south.
In Fig.~\ref{fig02}
we show the NIR continuum emission from UGC~1347 together with
the digitized sky survey V-band image in Fig.~\ref{fig03}.

\begin{figure*}
\vspace{-2cm}
\resizebox{12cm}{!}{\includegraphics{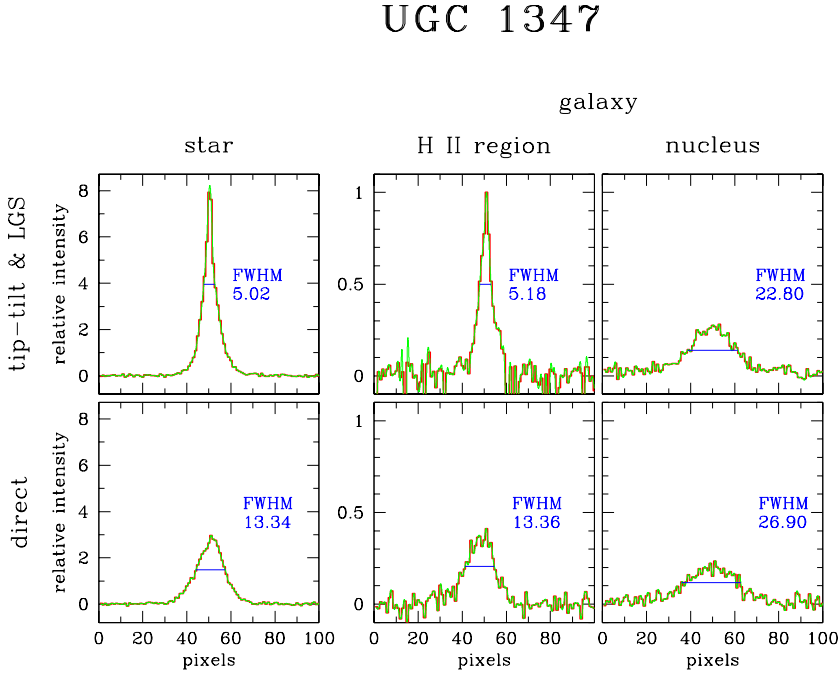}}
\hfill
\parbox[b]{55mm}{
\caption{E-W cuts through the direct and AO-corrected images of the
nucleus, the compact HII region at the tip of the bar,
and the tip-tilt reference star of UGC~1347. The FWHM of the images are given in pixels.
One pixel corresponds to 0.08$''$.}
\label{fig04}}
\end{figure*}

In Fig.~\ref{fig04} we show NIR intensity profile cuts of the
nucleus, the bright off-nuclear source, and the star PM~1111 in both
open loop and closed loop.
We have used the image of PPM~1111 as the point spread function and
deconvolved the NIR continuum image of UGC~1347 with a Lucy-Richardson
algorithm (Lucy \cite{Lucy}).
From the comparison of the images in Figures \ref{fig02} and \ref{fig03}
and the deconvolved image it is 
evident that the northern component coincides with the
nucleus of UGC~1347 and has an extent of about 1$''$ (corrected for the
FWHM of the PSF; see Tab.~\ref{ww01}) corresponding to a
diameter of about 310~pc.
The southern compact component is located at the southern tip of the
galaxy bar and is unresolved compared to the 0.40$''$ FWHM PSF.
We estimate an upper limit to its angular extent of
0.15$''$ corresponding to less than 45~pc.

The J, H, and K flux densities listed in Tab.~\ref{ww04} were measured in
4.8$''$, 3.6$''$, 2.4$''$, 1.2$''$ and 0.72$''$
diameter circular apertures centered on each component.

\begin{figure*}
\resizebox{12cm}{!}{\includegraphics{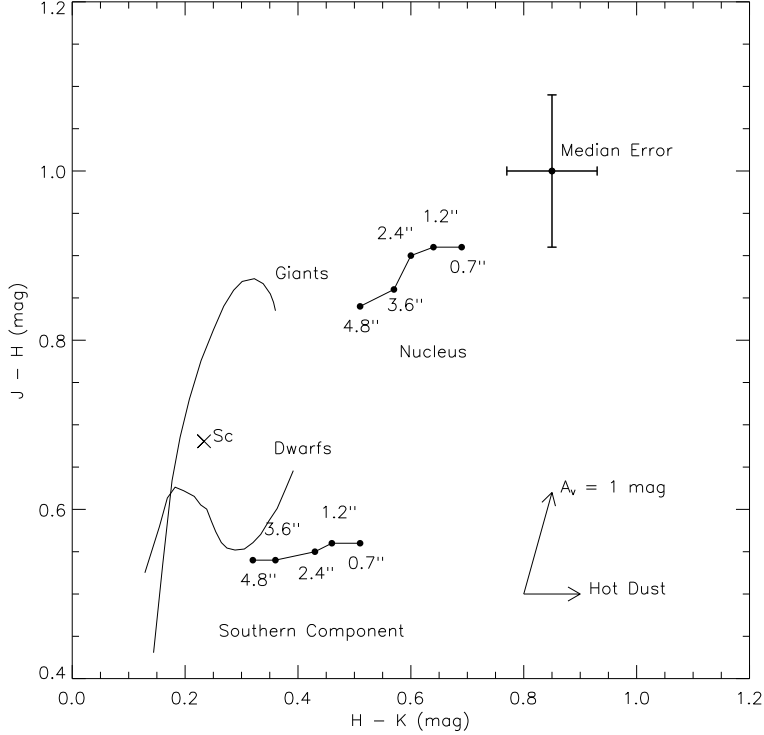}}
\hfill
\parbox[b]{55mm}{
\caption{ JHK two color diagram and the locations taken in
 by the two components. The JHK flux densities were measured in
 4.8$''$, 3.6$''$, 2.4$''$, 1.2$''$, 0.72$''$
 diameter circular apertures centered on each
 component. The graph indicates that the nuclear colors are in agreement
 with a stellar population reddened with an $A_{\rm V}$ of about 4 magnitudes.}
\label{fig05}}
\end{figure*}

In Fig.~\ref{fig05}
we show the locations of the multi-aperture data in the
J~$-$~H, H~$-$~K two color diagram.
The graph indicates that the nuclear colors are in agreement
with a stellar disk population reddened with an $A_{\rm V}$ of about 4 mag
corresponding to about 0.6 magnitudes of reddening in the K band. 
The value of $A_{\rm V}$ = 4~mag is large with respect to color variations due to 
differences in stellar population, age, and metallicity from a ``normal``
Sc disk population. Here we assumed a screen model for the extinction. 
In case of a mixed model the extinction may be even large.
In addition
to simple reddening there may also be some contribution from hot dust
to the nuclear NIR emission.
In the case of the southern component, however, the colors from a
normal stellar disk population are apparently more influenced by
additional emission from hot dust
and an extinction of $A_{\rm V} \le$ 2~mag
(corresponding to the arrows in the figure ledgend).
In both components the reddening and the influence from hot dust
emission increase with decreasing aperture size. This indicates
a decrease in dilution by a surrounding or underlying
stellar population unaffected by reddening.
The southern component is probably similar to the red knot found in
the nearby spiral NGC~7552 (Schinnerer et al. \cite{Schinnerer}) and is
likely region of recent active star formation in the disk.
However, its emission may very well be contaminated by red super-giants
(see results of our starburst analysis in section 4.2).
The contribution of hot dust in the nucleus may be
indicative for star formation activity there as well.

In order to analyze the NIR data in conjunction with other
data taken from the literature we used a starburst model as described
in section 4.


\small
\begin{table*}
\caption{\label{ww04} Photometric results on UGC~1347 and UGC~1344.}
\begin{center}
\begin{tabular}{cccccccccc}\hline \hline
\multicolumn{2}{r}{aperture diameter} & 4.8$''$& 3.6$''$& 2.4$''$& 1.2$''$& 0.72$''$\\
&  &     &     &     &     &      \\
\hline
{\bf UGC~1347}        &&&&&&\\
nucleus & J& \( 14.41\pm 0.08 \)& \( 14.84\pm 0.06 \)& \( 15.53\pm 0.05 \)& \( 16.85\pm 0.06 \)& \( 17.94\pm 0.06 \)\\
        & H& \( 13.57\pm 0.12 \)& \( 13.98\pm 0.09 \)& \( 14.63\pm 0.07 \)& \( 15.94\pm 0.07 \)& \( 17.03\pm 0.07 \)\\
        & K& \( 13.06\pm 0.01 \)& \( 13.41\pm 0.02 \)& \( 14.03\pm 0.01 \)& \( 15.30\pm 0.02 \)& \( 16.34\pm 0.04 \)\\
        & H $-$ K& \( 0.51\pm 0.12 \)& \( 0.57\pm 0.09 \)& \( 0.60\pm 0.07 \)& \( 0.64\pm 0.07 \)& \( 0.69\pm 0.08 \)\\
        & J $-$ H& \( 0.84\pm 0.14 \)& \( 0.86\pm 0.11 \)& \( 0.90\pm 0.09 \)& \( 0.91\pm 0.09 \)& \( 0.91\pm 0.09 \)\\
        &&&&&&\\
southern& J& \( 14.33\pm 0.07 \)& \( 14.65\pm 0.05 \)& \( 15.21\pm 0.04 \)& \( 16.43\pm 0.04 \)& \( 17.47\pm 0.03 \)\\
component
        & H& \( 13.79\pm 0.15 \)& \( 14.11\pm 0.10 \)& \( 14.66\pm 0.07 \)& \( 15.87\pm 0.06 \)& \( 16.91\pm 0.06 \)\\
        & K& \( 13.47\pm 0.01 \)& \( 13.75\pm 0.03 \)& \( 14.23\pm 0.01 \)& \( 15.41\pm 0.02 \)& \( 16.40\pm 0.05 \)\\
        & H $-$ K&\( 0.32\pm 0.15 \)&\( 0.36\pm 0.10 \)&\( 0.43\pm 0.07 \)&\( 0.46\pm 0.06 \)&\( 0.51\pm 0.08 \)\\
        & J $-$ H&\( 0.54\pm 0.17 \)& \( 0.54\pm 0.11 \)& \( 0.55\pm 0.08 \)& \( 0.56\pm 0.07 \)& \( 0.56\pm 0.07 \)\\
        &&&&&&\\
{\bf UGC~1344}        &&&&&&\\
nucleus & K& \( 12.11\pm 0.02 \)& \( 12.45\pm 0.03 \)& \( 12.96\pm 0.03 \)& \( 14.15\pm 0.04 \)& \( 15.12\pm 0.03 \)\\
\hline
\end{tabular}
\end{center}
\end{table*}
\normalsize

\subsubsection{The K-band luminosity}

Our K-band images allow us to calculate the nuclear
K-band
luminosity $L_{\rm K}$ and compare it to an estimate of
the K-band luminosity from the extended disk
and southern component
of UGC~1347.
The results are listed in Tab.~\ref{ww05} and Tab.~\ref{ww06}.

The K-band luminosity is calculated via
\begin{equation}
L_{\rm K} [L_{\odot}]=1.14 \times 10^4 \times D[{\rm Mpc}]^2 \times S_{\rm K} [{\rm mJy}]
\end{equation}
\noindent
where $D$ is the distance in Mpc and $S_{\rm K}$ is the 2.2 $\mu$m flux density
in mJy (Krabbe et al. \cite{Krabbe}).
We found K-band flux densities of 5.5~mJy for the extended nucleus and
3.7~mJy for the southern compact component
measured in 7.2$''$ diameter apertures.
In the previous section we have shown that the K-band flux density of
the southern component is clearly contaminated by emission of hot dust.
The J~$-$~H colors are less effected by hot dust emission
and correspond to that of an un-reddened stellar population.
We have therefore calculated the stellar K-band flux
density from the H-band magnitude using the standard spiral disk
colors with H~$-$~K = 0.21 (e.g. Frogel et al. \cite{Frogel}).
The K-band flux density of the southern component
corrected for the contribution of hot dust is then reduced to
about 3~mJy ($m_{\rm K}$ = 13.3~mag) resulting in a K-band luminosity of
1.5$\times$10$^8$~\solars.

In order to estimate the K-band luminosity of the disk
we used the total H-band
magnitude of 10.48 mag measured by Gavazzi et al. (\cite{Gavazzi1996b})
For a mean H~$-$~K color of 0.21 (Frogel et al. \cite{Frogel})
this results in a total K-band flux of the order of 49~mJy
corresponding to a total K-band luminosity of $L_{\rm K}$ = 2.1$\times$10$^9$
\solar.
Correcting for the contribution of the nucleus and the southern component
we obtain a K-band flux and luminosity for the disk of UGC~1347
of about 40~mJy and $L_{\rm K}$ = 1.6$\times$10$^9$~\solar.

\begin{table}
\caption{\label{ww05} Diagnostic parameters for UGC~1344 and
UGC~1347. $M_{\rm H_2}$ estimated from $L_{\rm FIR}$; $L_{K, {\rm
total}}$ estimated from H-band magnitude;
$M_{\rm total}$ estimated from the width of the HI-line.
For further information see text.}
\begin{center}
\begin{tabular}{lcccccc}\hline \hline
                          & UGC~1347                 &UGC 1344                     \\
                          &                          &                             \\ \hline
$M_{\rm H_2}$ [\solm]         & 3.2$\times$10$^9$        & $<$10$^9$                   \\
$M_{\rm HI}$[\solm]           & 3.2$\times$10$^9$        & $\le$3.3$\times$10$^8$      \\
$M_{\rm total}$ [\solm]       & (0.5-1)$\times$10$^{11}$ & (0.8-1.2)$\times$10$^{10}$  \\
$L_{\rm K}$ [\solar]            & 2.1$\times$10$^9$        & 3.7$\times$10$^9$           \\
$L_{\rm Lyc}$ [\solar]        & 5.5$\times$10$^8$        & $<$10$^8$                   \\
$L_{\rm FIR}$ [\solar]        & 1.18$\times$10$^{10}$    & $<$4$\times$10$^{9}$        \\
$\nu_{\rm SN}$ [yr$^{-1}$]   & 0.044                    & 0.012                       \\ \hline
$M_{\rm total}/L_{\rm K}$       & 24 -- 48     & 2.2 -- 3.3       \\
$L_{\rm bol}/L_{\rm Lyc}$       & 21           & --               \\
$L_{\rm bol}/L_{\rm K}$         &  5.6         & $<$1.1           \\
$L_{\rm K}/L_{\rm Lyc}$           &  3.8         & $>$37           \\
10$^{9}\nu_{\rm SN}/L_{\rm Lyc}$& 0.08         & $>$0.12          \\
10$^{12}\nu_{\rm SN}/L_{\rm bol}$& 3.7         & $>$3             \\
\hline 
\end{tabular}
\end{center}
\end{table}
\normalsize

\small
\begin{table}
\caption{\label{ww06} Diagnostic parameters for the UGC~1347
components. The data on the nucleus and the southern component have been obtained
in 7.2$''$ diameter apertures. $L_{\rm K}$ and $L_{\rm Lyc}$ for the disk do not
include the contributions from the nucleus and the southern component.
$L_{\rm FIR}$ for the disk has been separated from the nuclear
contribution on the assumption that the 100~$\mu$m flux density is
dominated by the disk and that a standard log($S_{60}/S_{100}$) ratio
can be applied for the disk.
The stellar $L_{\rm K}$ for the southern component has been calculated
from the H-band magnitude using the standard spiral disk colors with H~$-$~K
in order to avoid contamination of $L_{\rm K}$ by a contribution from hot dust.
$\nu_{\rm SN}$ has been derived on the basis of radio measurements which are
assumed to be associated with the nuclear region only.
For further information see text.}
\begin{center}
\begin{tabular}{lcccccc}\hline \hline
                          & Nucleus                  & Disk                        & Southern                  \\
                          &                          &                             & Component                 \\
                          &                          &                             &    \\ \hline
$L_{\rm K}$ [\solar]            & 3.7$\times$10$^8$        & 1.6$\times$10$^9$           & 1.5$\times$10$^8$ \\
$L_{\rm Lyc}$ [\solar]        & 9$\times$10$^7$        & 4.2$\times$10$^8$           & 4.2$\times$10$^7$ \\
$L_{\rm FIR}$ [\solar]        & 8.6$\times$10$^{9}$      & 3.2$\times$10$^{9}$         & -- \\
$\nu_{\rm SN}$ [yr$^{-1}$]   & 0.044                    & $<$0.002                    & -- \\ \hline
$L_{\rm bol}/L_{\rm K}$         & 23      &  2.0    & --  \\
$L_{\rm bol}/L_{\rm Lyc}$       & 95      &  7.6    & --  \\
$L_{\rm K}/L_{\rm Lyc}$           & 4.1     &  3.8    & 4 \\
10$^{9}\nu_{\rm SN}/L_{\rm Lyc}$& 0.49    & $<$0.005& --  \\
10$^{12}\nu_{\rm SN}/L_{\rm bol}$& 5.1    & $<$0.6  & --  \\
\hline 
\end{tabular}
\end{center}
\end{table}
\normalsize

\subsubsection{The Lyman continuum luminosity}

For the overall galaxy as well as for individual components an
estimate of the  H$\alpha$ luminosities have been obtained using the
continuum and H$\alpha$ line images kindly provided by
Amram et al. (\cite{Amram}; P. Amram  and M. Marcelin 1998, private
communication). Both the continuum and H$\alpha$ line images
were taken simultaneously with the same spectral resolution.
They have a field of view of $4.9' \times 4.9'$ and include all of
UGC~1347.
Main purpose of these observations was to derive the H$\alpha$ velocity
field. In order to determine flux densities we
used an estimate of the total H$\alpha$ continuum flux density
of UGC~1347 to calibrate the data.
The H$\alpha$ continuum flux density
was obtained via a linear interpolation between the
flux densities derived from the total
H-band and V-band magnitudes as given in Gavazzi \& Boselli
(\cite{Gavazzi1996a}). 
Although uncertain by probably 30\% this estimate allows us to further 
probe the consistency of the available data with our own
measurements and starburst analysis.
The corresponding calibration factor between the measured and calculated
H$\alpha$ continuum was applied to the  H$\alpha$ line data.
Finally the Lyman continuum luminosity $L_{\rm Lyc}$ then was
derived from the H$\alpha$ flux density $F_{\rm H \alpha}$ and
the source distance $D$ via:
\begin{equation}
L_{\rm Lyc} [L_{\odot}] =
5.6 \times 10^{17} \times F_{\rm H\alpha}[{\rm erg s}^{-1} {\rm cm}^{-2}]
\times D[{\rm Mpc}]^2
\end{equation}
(following Osterbrock \cite{Osterbrock}).
Without  extinction correction we obtain about
2.3$\times$10$^8$~\solarb for the overall Lyman continuum luminosity
of UGC~1347. 
In 7.2$''$ diameter apertures centered on the
nucleus and on the southern component we obtain approximately 10$^7$~\solarb
and 2$\times$10$^7$~\solarb, respectively.
In the following we assume the following values for the extinction, which are based on the
JHK measurements and (for the disk) on comparisons to other galaxies:
nucleus $A_{\rm V}$ = 3~mag, southern component $A_{\rm V}$ = 1~mag, and disk
$A_{\rm V} \le$ 1~mag.
Furthermore we use $A_{\rm H\alpha}=0.8 A_{\rm V}$ (Draine \cite{Draine}).
The corresponding extinction-corrected Lyman continuum luminosities
are given in Tab.~\ref{ww05} and Tab.~\ref{ww06}.

\subsubsection{The gas content and total mass}

The HI content of UGC~1347 was measured by Wilkerson (\cite{Wilkerson}) using the
Arecibo telescope with a 3.2$'$ beam. The distribution of atomic hydrogen in UGC~1347
was studied by Bravo-Alfaro (\cite{Bravo-Alfaro}) using the Westerbork
Synthesis Radio Telescope. The interferometric measurements
show that the HI gas extends well beyond the optical disk although the FWHM
of the distribution is in approximate agreement with the optical extent.
The HI (Bravo-Alfaro \cite{Bravo-Alfaro}) and H$\alpha$ (Amram et al. \cite{Amram}) rotation curves
are in good agreement.
The image suggests a slight HI line flux enhancement to the south-east
and extensions to the west and north.
Assuming a distance of 67~Mpc Wilkerson (\cite{Wilkerson}) obtained
3.5$\times$10$^9$~\solm of atomic hydrogen gas containing most
of the overall HI content of the galaxy.
Scaled to our adopted distance of 63~Mpc this results in
M$_{\rm HI}$ = 3.2$\times$10$^9$~\solm.
\\
No direct measurement of the
molecular gas mass in UGC~1347 is available. As an estimate we use the
IRAS far-infrared (FIR) flux densities and a dust temperature of 22~K
(see the following section)  and
the molecular hydrogen mass to L$_{\rm FIR}$ correlation by Young \&
Scoville (\cite{Young}). We estimate a molecular hydrogen mass of approximately
$M_{\rm H_2}$ = 3.2$\times$10$^9$~\solm. The resulting $M_{\rm H_2}/M_{\rm HI}$
ratio of 1 is then consistent with the value expected for late type spirals 
(Young \& Scoville \cite{Young}).
Although the determination of the molecular gas mass is uncertain, it is very unlikely that it is wrong by a large factor, say 10, since the age derived from
$(M_{\rm tot} - M_{\rm HI} - M_{\rm H_2}) / L_{\rm K}$ is fully consistent 
with the $L_{\rm bol} / L_{\rm Lyc}$ and $L_{\rm K} / L_{\rm Lyc}$ ratios (see section 4.2) 
for tis age.
We note that we implicitly assume a standard $N_{\rm H_2}/I({\rm CO})$ 
conversion factor. We also note, that  $M_{\rm H_2} / M_{\rm HI}$ is of
the order of 0.1 for optical selected galaxies, while $M_{\rm
H_2}/M_{\rm HI} \sim 1$
is found for infrared selected samples. This is consistent with
UGC~1347 being listed in the IRAS point source catalogue (see below).
\\
The measured HI line width of 144~km~s$^{-1}$ (Wilkerson \cite{Wilkerson})
and the H$\alpha$  velocity field corrected for
the inclination of $i=30^\circ$ (Amram et al. \cite{Amram}) indicates a full
velocity width covered by the rotation curve of $\Delta v_0 \approx 300$~km~s$^{-1}$.
Following Shostak (\cite{Shostak}) and Heckman et al. (\cite{Heckman})
this allows to estimate the total dynamical mass of UGC~1347
as $M_{\rm dyn}$ = (0.5--1.0)$\times$10$^{11}$~\solm. The resulting total gas to
dynamical mass ratio ($M_{\rm HI} + M_{\rm H_2}) / M_{\rm dyn}$ of about 5\% to 10\%
is in agreement with typical values found for spiral galaxies
(Shostak \cite{Shostak}).

\subsubsection{The FIR luminosity}

The FIR luminosity can be derived using the 60 $\mu$m and 100 $\mu$m IRAS flux
densities of $S_{60}$ = 1.40~Jy of $S_{100}$ = 3.84~Jy as
listed in the IRAS point source catalogue.
At both wavelengths UGC~1347 is fully contained in the large IRAS beams.
Following the formalism given by Lonsdale et
al. (\cite{Lonsdale}) and Fairclough (\cite{Fairclough}) we find a total
$L_{\rm FIR} = 1.18 \times 10^{10} h^{-2}$~\solarb.
From the 60 $\mu$m and 100 $\mu$m data we calculate a
dust color temperature of 22~K assuming an emissivity
proportional to $\lambda^{-1}$  (Hildebrand \cite{Hildebrand}) and a silicate to graphite
ratio of 7:3 (Whittet \cite{Whittet}).

We can also estimate how large the disk and nuclear contributions
to the FIR luminosity are. Here we assume that a dominant fraction
of the disk FIR
emission originates in diffuse interstellar dust and gas clouds
which are heated by the interstellar UV radiation field of the stellar
disk population. These clouds have
correspondingly large volume and disk area filling factors.
To obtain a first order estimate of the disk and nuclear contributions
we assume that for this population
of clouds the FIR flux densities  can be estimated
by adopting the relation found for ``cirrus'' emission in our Galaxy
(De Vries et al. \cite{De Vries}, Helou \cite{Helou}).
These relations have already successfully been applied to extragalactic
objects by Eckart et al. (\cite{Eckart1990}) for Centaurus~A and by
Jackson et al. (\cite{Jackson}) for NGC~2903.
Following De Vries et al. (\cite{De Vries}) the far-IR flux density
$S_{100}$ of the ``cirrus'' emission in Ursa Major can be obtained
via
\begin{equation}
S_{100} = a N_{\rm HI} + b I_{\,^{12}{\rm CO}(1-0)} +
S_{100, {\rm BG}},
\end{equation}
where $S_{100, {\rm BG}}$ is the flux density of the background
emission,
$N_{\rm HI}$ the HI column density, and $I_{\,^{12}{\rm CO}(1-0)}$ the integrated
$^{12}$CO(1$-$0) line flux.
The authors determine the constants $a$ and $b$ as
$a = (1.0 \pm 0.4) \times 10^{-20}$ MJy sr$^{-1}$ cm$^{2}$ and
$b = (1.0 \pm 0.5)$ MJy sr$^{-1}$ K$^{-1}$~km$^{-1}$ s.
Here we assume that IRAS point source data do not have to be corrected
for significant contributions of any background emission.
In order to obtain a lower limit to the FIR contribution of the disk
to the overall FIR luminosity we just calculate the contribution
expected from the atomic HI gas which is mostly distributed throughout
the disk.
The disk diameter of about 24.5~kpc and the adopted HI mass
of $M_{\rm HI}$ = 3.2$\times$10$^9$~\solm result in a 100 $\mu$m flux density
contribution of about 8.6 MJy sr$^{-1}$. Integrated over the disk
this gives a flux density of $S_{100}$ = 1.26 Jy.
For a total  molecular gas mass of 3.2$\times$10$^9$~\solm we find
a 100~$\mu$m flux density contribution of the order of 0.3~mJy.
The total disk flux density at 100 $\mu$m thus amounts to 1.56~Jy.
With a mean ratio between the 60 and 100 $\mu$m flux density
contribution for cirrus clouds of log($S_{60}/S_{100})=-0.65$
(Helou \cite{Helou}) the expected 60 $\mu$m flux density from the
disk is $S_{60}$ = 0.35~Jy.
These disk values can now be used to derive the nuclear FIR flux densities
and luminosity (Tab.~\ref{ww05} and Tab.~\ref{ww06}).
Here we assume that the contribution of the southern compact
component to the FIR luminosity is negligible because of the
small filling factor of the source in the IRAS beam.

Although the presented decomposition of the FIR flux densities is 
very indirect, the FIR luminosities of the nucleus and disk are consistent
with what one expects from the relation between the radio continuum and the 
FIR luminosity (Wunderlich \& Klein, \cite{Wunderlich}).

\subsubsection{The supernova rate}

Oly and Israel (\cite{Oly}) measured a 327~MHz flux density of UGC~1347 of
$S_{327~{\rm MHz}}$ = 23.6~mJy in a 55$''$ beam. The difference between
peak and integrated flux density is only $+$0.5~mJy. The disk size of
UGC~1347 is of the order of 90$''$ diameter. If the radio flux density
were be dominated by the central 10\% (30\%) of the disk this would
result in an approximately 3\% (30\%) deviation between the two
quantities.
In addition the source shows no clear indications for extended emission 
in the NRAO VLA Sky survey at 1.4~GHz (Condon et al. \cite{Condon1996})
with a beam size of 45$''$.
Therefore we assume that almost all
of the flux density can be attributed to the nuclear region
and that less than 1/20 of the radio emission originates in the disk of
UGC~1347.
Using a mean spectral index ($S\propto \nu^{-\alpha}$)
between 327~MHz and 1420/5000~MHz of $-0.71 \pm 0.05$, which the authors
obtained for a sample of 35 UGC galaxies, we estimate a 5~GHz flux
density of $S_{5~{\rm GHz}}$ = 3.41~mJy.
This value can be used to calculate the supernova rate $\nu_{\rm SN}$  via
\begin{equation}
\nu_{\rm SN} [{\rm yr}^{-1}]=3.1 \times 10^{-6} \times S_{5 {\rm GHz}}[{\rm mJy}] \times D[{\rm Mpc}]^2
\end{equation}
\noindent
(Condon \cite{Condon1992}).
For UGC~1347 we find a supernova rate of $\nu_{\rm SN}$ = 0.044~yr$^{-1}$.
This value for the nuclear region of UGC~1347
is of the same order as the estimated
overall supernova rate in the Milky Way of (0.025 $\pm$ 0.006)~yr$^{-1}$
(Tammann et al. \cite{Tammann}).
For the disk of UGC~1347 we adopt an upper limit of 
$\nu_{\rm SN}$ = 0.002~yr$^{-1}$.
This determination of the SNR assumes that there is no major contribution 
to the radio flux density by an active nucleus. In all star burst analyzes
were such a contribution could be excluded or was unlikely the derived SNR
have shown to be consistent with other measurements of the star formation
activity in the framework of the star burst model calculations.

\subsection{UGC~1344}

UGC~1344 is a SBc galaxy with an inclination of about $i=60^\circ$
(data in UGC catalogue and visual inspection) located at
R.A.(2000) = 01$^h$52$^m$34.8$^s$ and DEC(2000) = 36$^\circ$30$'$02$''$ approximately
21$'$ north of the center of Abell~262. Like UGC~1347, it is
well within the inner region in which the
largest amount of HI deficiency is observed.
A bright field star (GSC~2319-0343, $m_{\rm V}$ = 11.0~mag) is located about 23$''$
to the south.
In Tab.~\ref{ww05} we list all parameters and estimates for UGC~1344 that
were derived in a similar way as described above for UGC~1347.

\begin{figure}
\resizebox{\hsize}{!}{\includegraphics{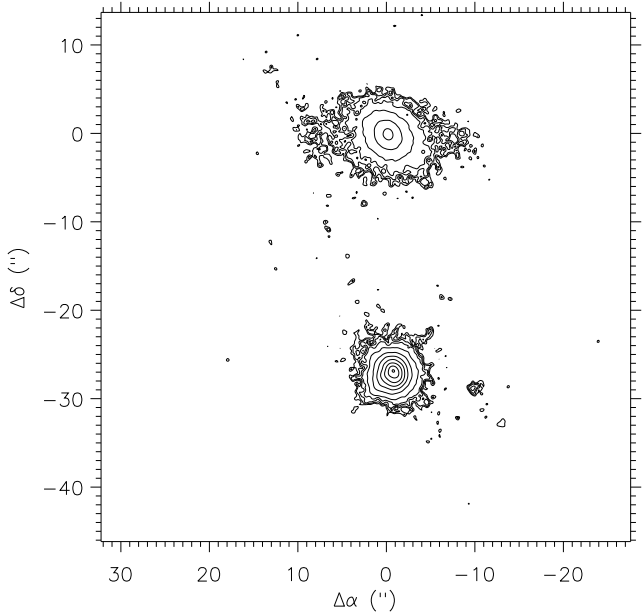}}
\caption{Closed-loop K-band image of UGC~1344 with a resolution of 0.4$''$.
The object to the south is the AO guide star.
The lowest contour level corresponds to 16.6~mag/arcsec$^2$.}
\label{fig06}
\end{figure}

\begin{figure}
\resizebox{\hsize}{!}{\includegraphics{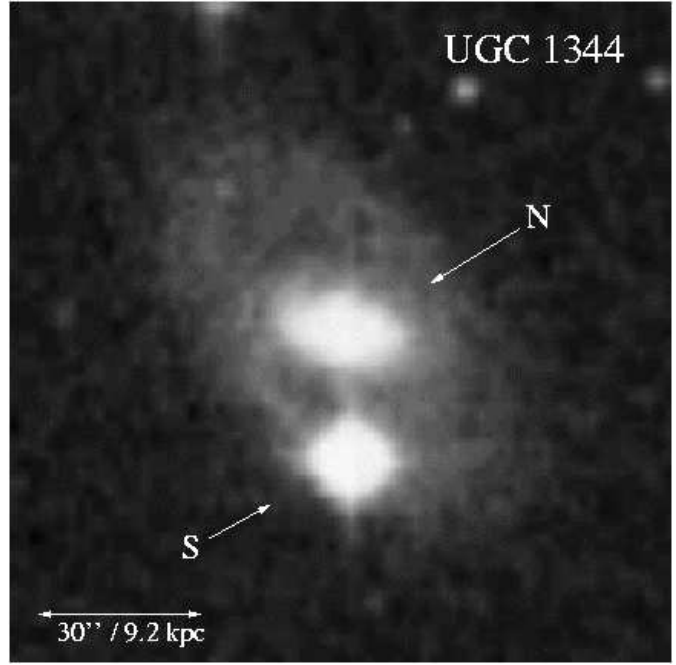}}
\caption{V-band image of UGC~1344 from the Palomar digitized sky survey.
The nucleus (N) and the guide star (S) are labeled.}
\label{fig07}
\end{figure}

\begin{figure*}
\vspace{-2cm}
\resizebox{12cm}{!}{\includegraphics{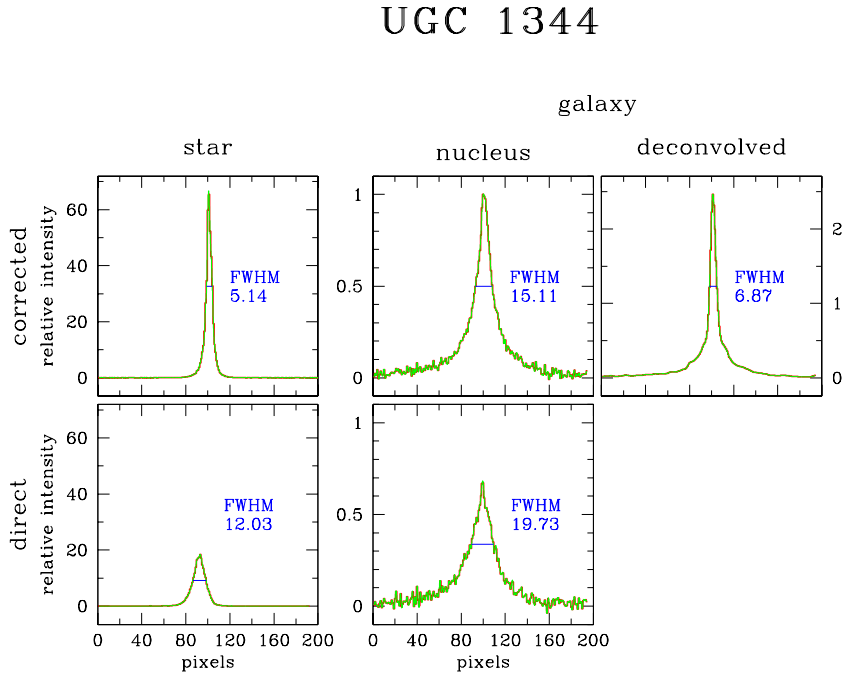}}
\hfill
\parbox[b]{55mm}{
\caption{E-W cuts through direct and AO-corrected images of
the nucleus and reference star of UGC~1344. A cut through a
LUCY deconvolved image of the nuclear component using the image of the
star as a PSF is also shown. The FWHM of the images are given in pixles.
One pixel corresponds to 0.08$''$.}
\label{fig08}}
\end{figure*}

Fig.~\ref{fig06}
shows the NIR continuum emission from UGC~1344 together with
the digitized sky survey V-band image in Fig.~\ref{fig07}.
We have used the image of the nearby reference star as the PSF and
deconvolved the NIR continuum image of UGC~1344 with a Lucy-Richardson
algorithm.
In Fig.~\ref{fig08}
we show intensity profile cuts through the
nuclear component of UGC~1344 and the reference star, for both open and
closed loop as well as deconvolved.

The K-band flux density distribution of UGC~1344 is very centrally peaked but
smoothly connects to the extended disk.
The central 3$''$ of the bulge contain about half of the K-band flux density
in a 10$''$ aperture.
For the nuclear component we measured a K-band flux density of 12~mJy
in a 7.2$''$ aperture. This is about a factor of 1.7 more than for UGC~1347.
As for UGC~1347 the K-band disk luminosity can be estimated from
the deep H-band image presented by Gavazzi et
al. (\cite{Gavazzi1996b}) as $m_{\rm H}$ = 9.81~mag.
For a mean H~$-$~K color of 0.21 (Frogel et al. \cite{Frogel})
this results in a total K-band flux of the order of 91~mJy.
Correcting for the contribution of the nucleus
we obtain a K-band flux and luminosity for the disk of UGC~1344
of about 81 mJy and $L_{\rm K}$ = 3.7$\times$10$^9$~\solar.


Amram et al. (\cite{Amram}) did not detect UGC~1344 in H$\alpha$ using a
similar integration time as for UGC~1347 (P. Amram  and M. Marcelin 1998,
private communication).
Based on the data
obtained for UGC~1347 we adopt here an upper limit for the
Lyman continuum luminosity of 10$^8$~\solarb for the entire galaxy.

The HI content of UGC~1344 has been studied by Wilkerson (\cite{Wilkerson}).
A weak line has been detected at a systemic velocity
of 4155~km~s$^{-1}$ and a width of 103~km~s$^{-1}$ (Wilkerson
\cite{Wilkerson}). Again we assume that the difference of approximately
550~km~s$^{-1}$ between the cluster
velocity  of 4704~km~s$^{-1}$ and the systemic velocity are due to the
motion of the galaxy within the cluster and adopt the same distance
of 63~Mpc as for UGC~1347.
From the HI detection Wilkerson (\cite{Wilkerson}) derived an upper limit of the
HI mass of 3.3$\times$10$^8$~\solm (scaled to the adopted distance
of 63~Mpc).

Oly and Israel (\cite{Oly}) measured the 327~MHz radio continuum
flux density of UGC~1347.
The difference between the integrated 327~MHz flux density of 6.2~mJy
and the peak flux density of 4.99~mJy in a $55'' \times 68''$ beam
indicates that not all of the radio emission can be associated with the
nuclear component but that at least 1~mJy is due to extended emission.
If the radio emission had a Gaussian distribution the
angular size would be of the order of about 30$''$ to 40$''$ suggesting that
this possible extended emission is distributed over the entire disk of
UGC~1344. 

From an inspection of the IRAS all sky survey we estimate an upper limit of
the flux density at a wavelength of 100 $\mu$m of 0.5~Jy.
Assuming a FIR spectrum similar to that of UGC~1347 this results in an
upper limit of the far-infrared luminosity of 4$\times$10$^{9}$~\solar.

\subsection{Near-infrared imaging of the sample}

In addition to the adaptive optics observations of UGC~1344 and UGC~1347 we
took seeing-limited images of additional 9 spiral galaxies (a total of 11)
in the Abell~262 cluster and 15 spiral galaxies in the Abell~1367 cluster.
They were selected according to their HI deficiency and separation from the
cluster center as given by Giovanelli et al. (\cite{Giovanelli1982})
and listed in Tab.~\ref{ww07} and Tab.~\ref{ww08}.

\begin{figure*}
\resizebox{12cm}{!}{\includegraphics{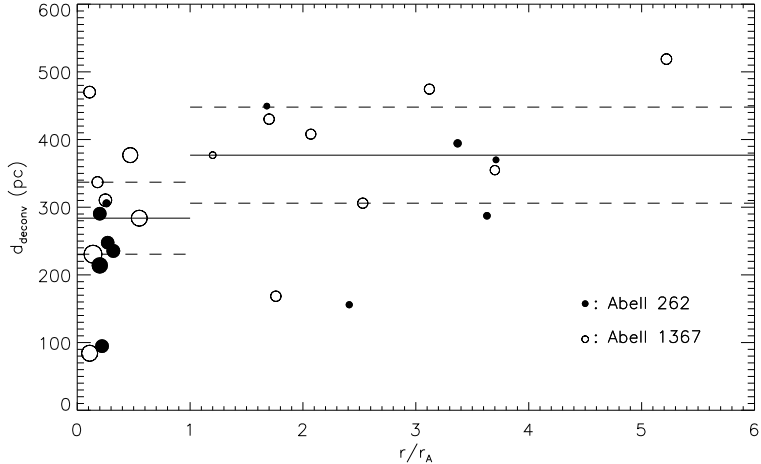}}
\hfill
\parbox[b]{55mm}{
\caption{Deconvolved nuclear source sizes in Abell 262 and Abell 1367 against
the distance from the cluster center (inside and outside of the Abell
radius). The size of the data points is
proportional to the HI deficiency of each galaxy.}
\label{fig09}}
\end{figure*}

A sample of 6 galaxies in Abell~262 and 7 galaxies in Abell~1367
was selected from the central parts of the clusters.
In both cases they have
separations from the cluster center less than 0.55 Abell radii
($r_A$ = 1.75$^\circ$ for Abell~262 and $r_A$ = 1.40$^\circ$ for Abell~1367)
and HI deficiencies (as defined in Giovanelli et al. \cite{Giovanelli1982})
ranging between 0.06 and $>$1.18.
The only exception is UGC~1347 with a deficiency of $-0.07$ which is
HI rich for a galaxy within the Abell radius.
A second sample of 5 galaxies in Abell~262 and 8 galaxies in Abell~1367
was selected from
the outer cluster regions. Here the separations from the cluster
center range from 1.2 to 5.2 Abell radii and the HI deficiencies
range from 0.03 to 0.6.
Our K-band images of these sources contain in almost all cases reference stars
that allowed us to accurately estimate the seeing.
If no star was contained in the image we used the stars in adjacent
exposures as a reference.
 From radial averages
centered on the galaxy nuclei and on the stars we extracted the sizes as FHWM
values (as listed in Tab.~\ref{ww07} and Tab.~\ref{ww08}) and deconvolved the measurements on the
galaxies with the stellar data via quadratic subtraction
assuming Gaussian flux distributions of the sources.
A comparison to our AO results on UGC~1347 and UGC~1344 shows
that this procedure gives reliable estimates of the bulge sizes.
 In Fig.~\ref{fig09}
we plot the deconvolved nuclear source sizes against the
 distance from the cluster center.  
To combine the data sets
 we scaled the results according to the mean
 radial velocities of the two galaxy clusters and corrected the Abell~1367
 data to the distance of the Abell~262 cluster.
 For the galaxies
 close to the cluster center we find a median nuclear FWHM
 and median deviation from that value of 0.77$''$ $\pm$ 0.07$''$.
 For the outer sources the result is 1.10$''$ $\pm$ 0.18$''$.
 The difference between the two median values is 2.6 times the mean of the
 two median deviations.
 A Kolmogorov-Smirnov test shows that the two distributions are different
 at the 85\% level.
 This result provides evidence that the K-band flux density distribution
 of the galaxy bulges in the inner part of the cluster are systematically
 smaller than those in the outer part.
This result is discussed in section 4.3.

\section{Discussion}

The data on UGC~1344 and UGC~1347
indicate ongoing or recent nuclear star formation.
The implications from that provide a framework in which the
properties of our Abell~262 and Abell~1367 sample can be explained as well.

\subsection{The starburst model}

To derive the properties of a starburst from the observed continuum and
line intensities we have used
the starburst code STARS.
This model has been successfully applied to NGC 1808 (Krabbe et al. \cite{Krabbe},
Tacconi-Garman et al. \cite{Tacconi-Garman}),
NGC~7469 (Genzel et al. \cite{Genzel}), NGC~6764 (Eckart et al. \cite{Eckart1996}) and NGC~7552
(Schinnerer et al. \cite{Schinnerer}). A description of the model can be found in the appendices of Krabbe
et al. (\cite{Krabbe}) and Schinnerer et al. (\cite{Schinnerer}).
The model is similar to other stellar population synthesis models
(Larson \& Tinsley \cite{Larson}, Rieke et al. \cite{Rieke1980}, Gehrz
et al. \cite{Gehrz},
Mas-Hesse \& Kunth \cite{Mas-Hesse}, Rieke et al. \cite{Rieke1993},
Doyon et al. \cite{Doyon}) and includes the most recent stellar evolution
tracks (Schaerer et al. \cite{Schaerer}, Meynet et al. \cite{Meynet}).

We assume power-law initial mass functions (IMFs) which vary as
$M^{-\alpha}$ between a lower and upper mass cut-off, 
$M_l$ = 1~\solm and $M_u$ = 100~\solm,
with an index $\alpha$ = 2.35 (Leitherer \cite{Leitherer}, Salpeter et al. \cite{Salpeter}).
STARS has
as output observable parameters such as the bolometric luminosity
$L_{\rm bol}$, 
the K band luminosity $L_{\rm K}$, the Lyman continuum luminosity $L_{\rm Lyc}$ and the
supernova rate $\nu_{\rm SN}$, as well as the diagnostic ratios between these
quantities.
The adopted values for $L_{\rm bol}$, $L_{\rm Lyc}$, $L_{\rm K}$, and $\nu_{\rm SN}$
have been derived from observed properties in the previous sections.
All relevant quantities as well as the diagnostic ratios that
can be calculated from them for the present analysis
are listed in Tab.~\ref{ww05} and Tab.~\ref{ww06}.
All the ratios are measures of the time evolution and the shape of the
IMF, with slightly different dependencies on $\alpha$ and $M_u$.

\begin{table*}
\caption{\label{ww07} The sample of spiral galaxies in
Abell~262. Sizes $d_{\rm deconv}$ of the galaxy nuclei extracted  
from seeing-limited images in K-band by quadratic deconvolution. Objects also observed in closed loop operation are indicated. 
The sizes evaluated from the
uncorrected images agree well with the corresponding results achieved from
the tip-tilt or fully AO corrected images, respectively.}
\begin{center}
\begin{tabular}{cccccccccc}\hline \hline
Galaxy& $r$/$r_A$ &HI-       &$d_{\rm core}$($''$)& $d_{\rm star}$($''$)& d$_{\rm deconv}$($''$)& Comment\\
      &           &Deficiency&              &               &                & \\
\hline
\object{UGC~1045} & 3.71 & -0.26   & 1.36 & 0.63 & 1.21 & \\
\object{UGC~~909} & 3.63 & -0.13   & 1.13 & 0.63 & 0.94 & \\
\object{UGC~1125} & 2.41 & -0.20   & 0.80 & 0.62 & 0.51 & \\
\object{UGC~1069} & 3.37 & -0.03   & 1.42 & 0.59 & 1.29 & \\
\object{UGC~1178} & 1.68 & -0.27   & 1.58 & 0.59 & 1.47 & tip-tilt\\
\object{UGC~1366} & 0.32 & $>$0.81 & 0.95 & 0.55 & 0.77 & \\
\object{UGC~1307} & 0.27 & $>$0.77 & 0.99 & 0.57 & 0.81 & \\
UGC~1347 & 0.26 & -0.07   & 1.24 & 0.73 & 1.00 & AO\\
\object{UGC~1338} & 0.22 & $>$0.79 & 0.81 & 0.59 & 0.31 & tip-tilt\\
UGC~1344 & 0.20 & $>$0.78 & 1.08 & 0.51 & 0.95 & AO\\
\object{UGC~1350} & 0.20 & $>$1.18 & 0.91 & 0.58 & 0.70 & \\
\hline
\end{tabular}
\end{center}
\end{table*}
\normalsize

\small
\begin{table*}
\caption{\label{ww08} The sample of spiral galaxies in Abell~1367. K-band sizes $d_{\rm deconv}$ of the galaxy nuclei extracted 
from seeing-limited images by quadratic deconvolution.}
\begin{center}
\begin{tabular}{cccccccccc}\hline \hline
Galaxy& $r$/$r_A$ &HI-       & $d_{\rm core}$($''$)& $d_{\rm star}$($''$)& $d_{\rm deconv}$($''$)\\
      &           &Deficiency&              &               &                \\
\hline
\object{UGC~7040} & 5.22 &  -0.01 &  1.26 & 0.47 & 1.17 \\
\object{UGC~7087} & 3.70 &  -0.20 &  0.94 & 0.48 & 0.80 \\
\object{UGC~6483} & 3.12 &  -0.08 &  1.26 & 0.66 & 1.07 \\
\object{UGC~6891} & 2.53 &  -0.06 &  0.84 & 0.48 & 0.69 \\
\object{UGC~6693} & 2.07 &  -0.09 &  1.02 & 0.44 & 0.92 \\
\object{UGC~6876} & 1.76 &  -0.04 &  0.90 & 0.82 & 0.38 \\
\object{UGC~6863} & 1.70 &  -0.04 &  1.14 & 0.60 & 0.97 \\
\object{UGC~6583} & 1.20 &  -0.60 &  1.12 & 0.72 & 0.85 \\
\object{UGC~6746} & 0.55 &$>$0.79 &  0.88 & 0.60 & 0.64 \\
\object{UGC~6663} & 0.47 &$>$0.65 &  1.01 & 0.54 & 0.85 \\
\object{UGC~6719} & 0.25 &   0.28 &  0.95 & 0.64 & 0.70 \\
\object{UGC~6702} & 0.18 &   0.06 &  1.01 & 0.66 & 0.76 \\
\object{UGC~6718} & 0.14 &   1.09 &  1.02 & 0.88 & 0.52 \\
\object{UGC~6688} & 0.11 &$>$0.80 &  0.53 & 0.49 & 0.19 \\
\object{UGC~6697} & 0.11 &   0.16 &  1.25 & 0.66 & 1.06 \\
\hline
\end{tabular}
\end{center}
\end{table*}
\normalsize

\subsection{Nuclear star formation in UGC~1347 and UGC~1344}

{\bf UGC~1347: }
The diagnostic ratios (Tab.~\ref{ww05} and \ref{ww06})
and the framework of
the starburst model now allow us to discuss the star formation
for the whole of UGC~1347 as well as for
the nucleus, the southern component and the disk.
In summary the overall disk data of UGC~1347 are consistent with a high age
and constant star formation, whereas the data for the nucleus and
the southern compact component indicate
more recent or ongoing star formation activity.

\begin{figure*}
\resizebox{12cm}{!}{\includegraphics{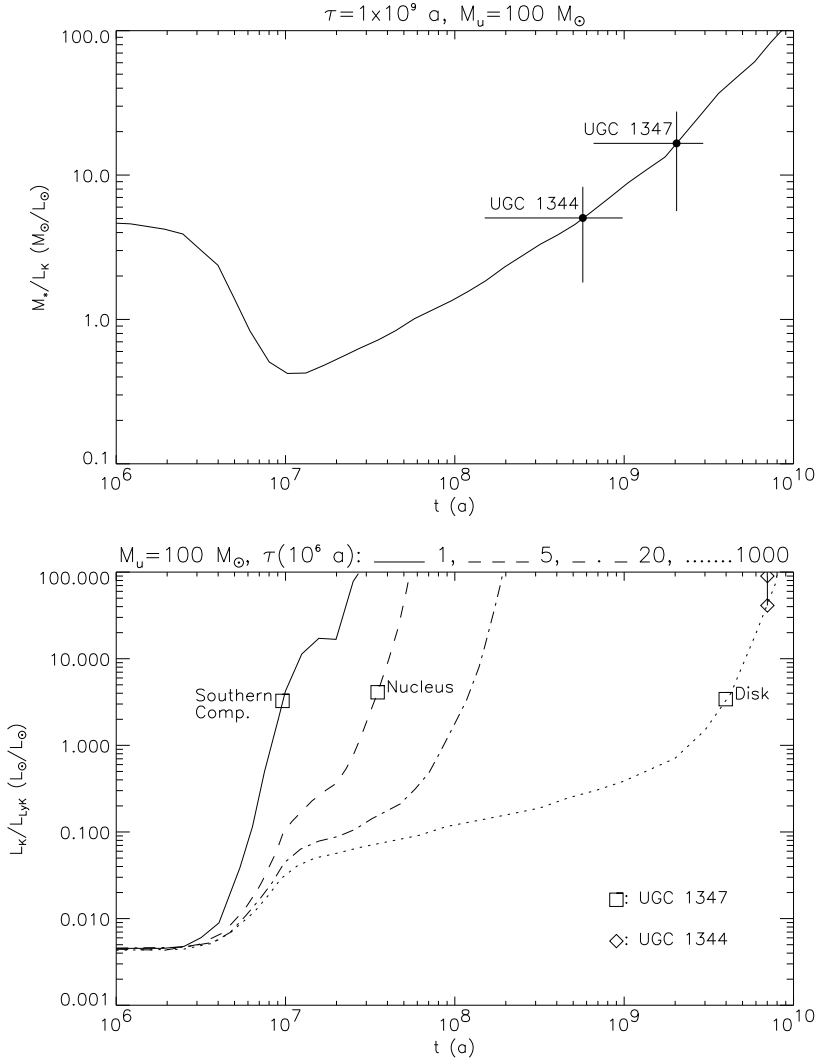}}
\hfill
\parbox[b]{55mm}{
\caption{Results from the starburst analysis: log($M_{\star}/L_{\rm K}$) (top) and
log($L_{\rm K}/L_{\rm Lyc}$) (bottom) both as a function of starburst age.
The ages for the overall galaxy derived from the mass-to-light ratio
are consistent in each case with the disk population ages derived from
the analysis of the luminosity ratios. For the discretization in the
starburst decay time constant the $L_{\rm K}/L_{\rm Lyc}$ data symbols display
the individual ages consistent with the
available $L_{\rm bol}/L_{\rm Lyc}$ and $\nu_{SN}/L_{\rm bol}$ ratios (model curves not
shown here).}
\label{fig10}}
\end{figure*}

{\em Disk:}
The $M_{\rm tot}$/ $L_{\rm K}$ ratio of 22 to 43 that we obtained for the
overall galaxy indicates an age of the stellar population in the disk
of the order of several 10$^9$ to 10$^{10}$ years (see Fig.~\ref{fig10}).
This is also supported by the ratios
$L_{\rm bol} / L_{\rm Lyc}$ = 6.4
and
$L_{\rm K}/L_{\rm Lyc}$ = 3.2
combined with
10$^{9}\nu_{\rm SN}/L_{\rm Lyc} < 0.004$
that results from the upper limit on the extended radio flux.
\\
{\em Nucleus:}
The diagnostic  ratios
$L_{\rm bol}/L_{\rm Lyc}$ = 53,
$L_{\rm K}/L_{\rm Lyc}$ = 2.3
combined with
10$^{9}\nu_{\rm SN}/L_{\rm Lyc}$ = 0.27
indicate that the nuclear light is currently
dominated by a decaying star formation event
that happened a few times 10$^7$ years ago. This assumes that the
upper mass cutoff  is 100~\solm and that the star burst itself
did not last longer than a few times 10$^6$ years (see Fig.~\ref{fig10}).
The initial star formation rate was 18~\solm~yr$^{-1}$ and the
current star formation rate would then be of the order of 0.04~\solm~yr$^{-1}$.

{\em The southern component:}
The southern component and the nucleus belong to the brightest regions
in H$\alpha$ line emission.
The only diagnostic ratio that we can calculate is that between the
K-band luminosity corrected for the contribution of hot dust
and the Lyman continuum luminosity. We obtain a value of
$L_{\rm K}/L_{\rm Lyc}$ = 4. In addition the compact source of hot dust
revealed by  our AO measurements and its location at the tip of
the bar suggest that the star formation activity in that
region may be as high and recent as we find it for the nucleus (see Fig.~\ref{fig10}).

{\bf UGC~1344: }
A ratio of
$M_{\rm tot}/L_{\rm K}$ = 2.2 to 3.3  indicates an age of 10$^9$ years (see
Fig.~\ref{fig10}).
This is supported by the lower limit $L_{\rm K}/L_{\rm Lyc} \ge 37$.
The fact that the far-infrared luminosity is lower than that of
UGC~1347 may imply that the molecular gas mass is below 10$^8$~\solm.
In addition the the system is deficient in neutral hydrogen as well.
This indicates that UGC~1344 is deficient overall in
fuel for star formation.
The fact that UGC~1344 shows only a weak and narrow HI line may indicate
that the HI has largely been stripped and the HI line width
cannot necessarily be taken as a measure for the total dynamical mass.
If the dynamical mass has been underestimated by  a factor of 10
the log($M_{\rm tot}/L_{\rm K}$) ratio will be similar to the one obtained for UGC~1347
and the age of the dominant stellar population in the disk
is then most likely also of the order of several 10$^9$ to 10$^{10}$ years.

\subsection{Nuclear star formation and bulge sizes in the sample}

The detailed discussion of the data available for UGC~1344 and UGC~1347
has shown that there is strong evidence for nuclear star formation
activity and that not all of the K-band nuclear flux density can be
explained by the presence of an old stellar population alone.
Additional K-band flux density may originate from hot dust
(see Fig.~\ref{fig05})
or a population of supergiants or AGB stars,
both of which are indicative for recent or ongoing star formation activity.
The spatial distribution of the sources responsible for additional K-band flux
density  may be looked upon  as independent of the distribution of
the old stellar population forming the bulge of the galaxies.
The relative flux density contribution of the additional sources will
then have an influence on the measured size of the bulge.

That star formation activity is an important quantity for the appearance and
classification of a galaxy has also been pointed out by Kennicutt et
al. (\cite{Kennicutt}). They have combined H$\alpha$ and UBV
measurements of 210 nearby Sa-Irr galaxies
with new photometric synthesis models to reanalyze past and future
star formation timescales in the disks.
The authors find that the pronounced change in the photometric properties of
spiral galaxies along the Hubble sequence is predominantly due to changes
in the star formation histories of disks, and only secondarily to changes
in the bulge/disk ratio.

It is well known that there is a strong morphological
segregation in clusters of galaxies, with most of the ellipticals
in the center of the cluster while the spirals are more dispersed.
Also early-type spirals seem more concentrated than late-types.
Although Abell~262 is an extremly spiral rich cluster this segregation
will most likely also effect the bulge to disk ratio as a function of
location in the cluster. It is, however, not self-evident that this
effect has any influence on the measured bulge sizes. In the following
we will address this problem.

Our finding can be discussed in the framework of recent
investigations of the bulge-to-disk luminosity ratio.
In a sample of 3114 galaxies Solanes et al. (\cite{Solanes}) analyzed
the luminosity of bulge and disk
components of disk galaxies and their possible correlations with
morphological type and local density.
Independently of the local environment no evidence is
found for any bulge segregation among disk galaxies. Instead they find
that disks appear to be less luminous with increasing local density.
They find that the absolute brightness difference $M$(bulge)$-M$(total)
corresponds to about 3 for Sc galaxies, 2 for Sb and 1 for Sa and S0 galaxies.
A similar trend is also observed in the dependency of the
near-infrared concentration index $C_{31}$
(defined in Gavazzi et al. \cite{Gavazzi1990}) on the absolute H-band
magnitude in a sample of 297 galaxies investigated by Gavazzi et al.
(\cite{Gavazzi1996b}). In this sample the bulge-to-disk ratio systematically
increases with decreasing H-band luminosity. The morphological
segregation reported by Solanes et al. (\cite{Solanes}) is (if at
all) only weakly indicated.

Both quantities, $M$(bulge)$-M$(total) and $C_{31}$, are concentration
parameters that simply describe the bulge versus disk brightness.
In a scenario in which bulge components of identical brightness are
located in the centers of disks of varying brightness the FWHM of the NIR
light,
measured with respect to the combined peak of the disk and bulge component
will result in smaller values for lower disk luminosities.
Therefore the observed disk luminosity antisegregation
could in principle
be responsible for our observed tendency that the FWHM
of galaxies within the Abell radii of the Abell~262 and Abell~1367
clusters are smaller than those outside the Abell radii.

However, it is also possible
that a luminosity antisegregation as observed
for the disks  is also present for the bulges, but is just compensated
for by the additional contribution to the bulge luminosity due to
enhanced star formation triggered by the effects of a higher
cluster density environment.

Theoretical studies (Moore et al. \cite{Moore}, Fujita \cite{Fujita}) show that the
velocity perturbation induced by a single high-speed encounter is
in most cases too small to affect the star formation rate of a disk galaxy.
However, several successive high-speed encounters between galaxies
(galaxy harassment) may lead to gas inflow and strong star
formation activity (Fujita \cite{Fujita}).
This picture is consistent with the cluster crossing and star formation
time scales.
From the sample of 84 Abell~262 cluster members listed in Giovanelli
\& Haynes (\cite{Giovanelli1985}) we derive a velocity dispersion of 750~km~s$^{-1}$.
The Abell radius of 1.75$^\circ$ then indicates that several 10$^9$
years are required to cross a significant fraction of the central
part of the cluster.
However, the time scale for formation of supergiants plus their life time
amounts to only several 10$^7$ years.
This indicates that in gas rich spiral galaxies star formation
can easily be triggered via galaxy-galaxy interactions while
passing through the central part of the cluster.

\subsection{Star formation in the high redshift clusters}

\begin{figure}
\resizebox{\hsize}{!}{\includegraphics{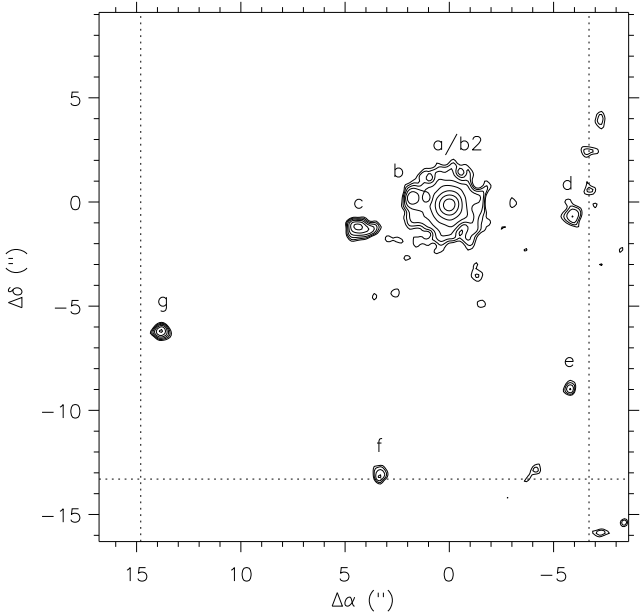}}
\caption{AO-corrected K$'$-band image of the J1836.3CR field with a
resolution of 0.15$''$. The lowest contour level
corresponds to 19.7~mag/arcsec$^2$; contour spacing is
0.5~mag/arcsec$^2$. Within the central dotted area the mosaic provides
the highest signal-to-noise ratio. The labeled sources are also
detected in J and H band.} 
\label{fig11}
\end{figure}

\begin{figure}
\resizebox{\hsize}{!}{\includegraphics{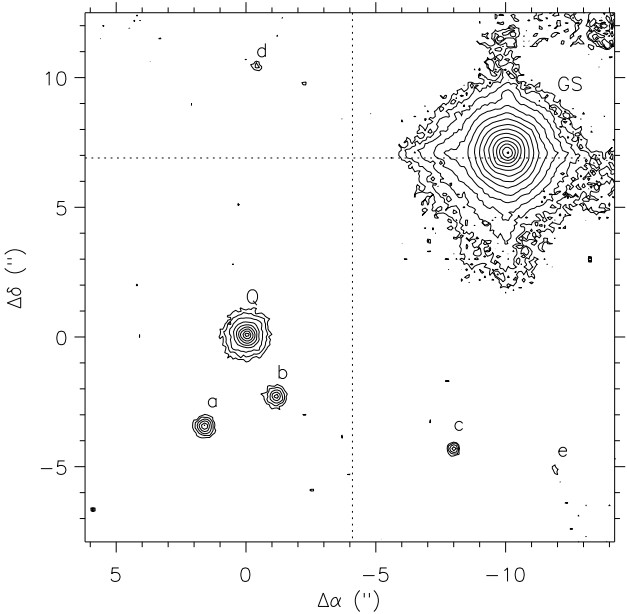}}
\caption{AO-corrected K$'$-band image of the PKS~0743-006 field with a
resolution of 0.15$''$. Within the southeastern dotted area the mosaic
provides the highest signal-to-noise ratio. The lowest contour level
corresponds to 18.5~mag/arcsec$^2$; contour spacing is
0.6~mag/arcsec$^2$.
The sources $a - d$ also appear in J- and H-band.}
\label{fig12}
\end{figure}

In Fig.~\ref{fig11}
 we show a K$'$-band image of the J1836.3CR field and
in Fig.~\ref{fig12}
 a K$'$-band image of the PKS~0743-006 field.

\begin{figure*}
\resizebox{12cm}{!}{\includegraphics{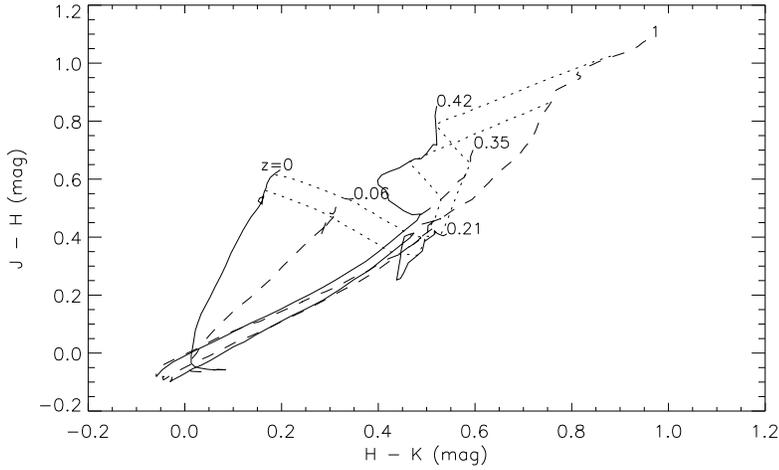}}
\hfill
\parbox[b]{55mm}{
\caption{Model values for the JHK flux densities calculated using
the GISSEL stellar population models (Bruzual \& Charlot
1993). Different redshifts are indicated. Every
evolution track starts (at the lower left corner) at time zero. The
dotted lines mark the location of a $10\times10^9$ yr and
$20\times10^9$ yr old population, respectively.}
\label{fig13}}
\end{figure*}

\begin{figure*}
\resizebox{12cm}{!}{\includegraphics{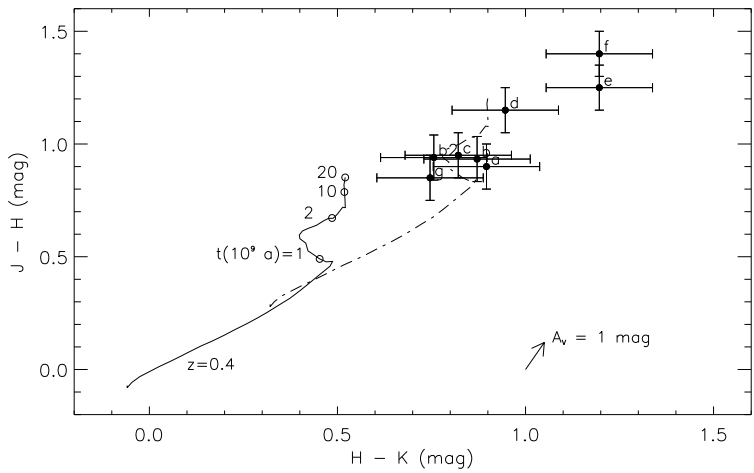}}
\hfill
\parbox[b]{55mm}{
\caption{GISSEL stellar population models for J1836.3CR at $z=0.414$. The data
points are best explained by a reddening of $A_{\rm V} = 3$~mag (dash-dotted curve).
Sources $e$ and $f$ are probably background galaxies.}
\label{fig14}}
\end{figure*}

\begin{figure*}
\resizebox{12cm}{!}{\includegraphics{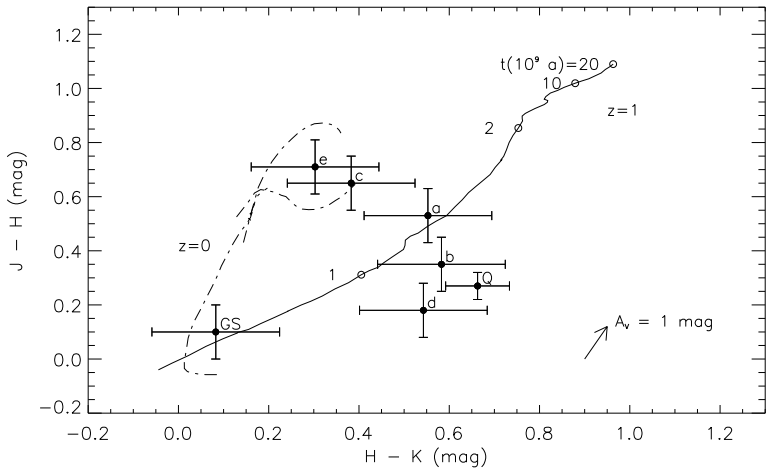}}
\hfill
\parbox[b]{55mm}{
\caption{GISSEL stellar population models for PKS~0743-006 at $z=0.994$. Drawn
are the unreddened evolution tracks for $z=0$ and $z=1$, respectively.}
\label{fig15}}
\end{figure*}

Model values for the JHK flux densities were calculated using the
GISSEL stellar population models (Bruzual \& Charlot \cite{Bruzual}),
and compared with the measurements for each field and appropriate
redshift. 
For better comparison we only show the tracks
in Fig.~\ref{fig13}, and then superimpose the
observed data in Fig.~\ref{fig14} and Fig.~\ref{fig15}.

All the models have been calculated for a passively evolving
population after a 1~Gyr starburst.
The Padova initial spectral energy distribution is used
with a Salpeter IMF. Continued star formation,
renewed star formation or initial mass functions truncated at their
high mass end all tend to move galaxies towards the starts of the tracks.
Since the SHARP~II+ data were taken in the K$'$ filter we transformed the
K$'$ magnitudes to K band values using the following relation:
${\rm K} = {\rm K}' + 0.2 \times ({\rm H}-{\rm K})$ (Wainscoat and Cowie \cite{Wainscoat}).
For the color correction we used the mean observed H~$-$~K colors of those
objects which we could identify as being extragalactic and close to the
redshift of the corresponding cluster.
For  the  PKS~0743-006 cluster these mean colors are
H~$-$~K $\sim 0.55$ and for the J1836.3CR cluster we find H~$-$~K $\sim 0.69$.
In Tab.~\ref{ww09} and Tab.~\ref{ww10} we give the magnitudes
 and flux densities and JHK colors
for the objects in our field of view towards both clusters.

In all cases the flux densities were derived from sky-subtracted images, 
taking into account possible contaminations by neighboring sources.
The mean apertures for J1836.3CR and PKS~0743 were 
3.95$''$(K$'$), 3.80$''$(H), 3.62$''$(J), and 
4.13$''$(K$'$), 3.60$''$(H), 3.33$''$(J), respectively.

{\bf J1836.3CR:} In Fig.~\ref{fig14}
we show the colors of all sources in our
field of view towards the J1836.3CR cluster at $z=0.414$.
All galaxies are located at the tip of the evolutionary track for a
passively evolving galaxy at that redshift. This is consistent with
an age of 10 Gyrs (or more) and some intrinsic reddening with a mean
value of the order of $A_{\rm V}$ = 2--3~mag.

\small
\begin{table*}
\caption{\label{ww09} Photometry and colors of sources in
J1836.3CR. The photometric quality of the data is of the order of
0.10$^m$ in the K- and H-band and 0.15$^m$ in the J-band. All sources can be identified as galaxies (G).
}
\begin{center}
\begin{tabular}{cccccccccc}\hline \hline
source&  ident. &  $J$   &  $H$   &  $K$   & $J-H$ & $H-K$ \\ \hline
a  & G &17.10 & 16.20 & 15.30 & 0.90 & 0.90\\
b  & G &20.23 & 19.30 & 18.43 & 0.93 & 0.87\\
b2 & G &19.59 & 18.64 & 17.89 & 0.94 & 0.77\\
c  & G &19.02 & 18.08 & 17.25 & 0.95 & 0.82\\
d  & G &20.85 & 19.70 & 18.75 & 1.15 & 0.95\\
e  & G &20.35 & 19.10 & 17.90 & 1.25 & 1.20\\
f  & G &20.20 & 18.80 & 17.60 & 1.40 & 1.20\\
g  & G &19.10 & 18.25 & 17.50 & 0.85 & 0.75\\
\hline 
\end{tabular}
\end{center}
\end{table*}
\normalsize
\small
\begin{table*}
\caption{\label{ww10} Photometry and colors of sources in
 PKS~0743-006. The photometric quality of the data is of the order of 0.10$^m$ in the K- 
and H-band and 0.15$^m$ in the J-band. Three sources in the vicinity of the quasar can be
 identified as galaxies (G). 
 }
\begin{center}
\begin{tabular}{cccccccccc}\hline \hline
source & ident.     &  J   &  H   &  K   & J $-$ H & H $-$ K \\ \hline
GS     & guide star & 11.05 & 10.95 & 10.90 & $-$0.05 &  $-$0.05 \\
Q      & quasar     & 15.80 & 15.50 & 14.90 &  0.30 &   0.60 \\
a      &   G        & 17.60 & 17.10 & 16.60 &  0.50 &   0.50 \\
b      &   G        & 17.70 & 17.40 & 16.80 &  0.40 &   0.50 \\
c      & star       & 18.00 & 17.80 & 16.60 &  0.70 &   0.10 \\
d      &   G        & 20.50 & 20.30 & 19.80 &  0.20 &   0.50 \\
e      & star       & 20.30 & 20.10 & 19.40 &  0.70 &   0.30 \\

\hline
\end{tabular}
\end{center}
\end{table*}
\normalsize
Source $a$ is the brightest galaxy in our field of view of the  J1836.3CR
cluster. For this source our data allow us for the first time to
obtain reliable color information close to the diffraction limit of the
3.6~m telescope. For this purpose the three 0.10$''$ pixel scale maps
were first deconvolved with the PSF and then reconvolved with a
Gaussian clean beam to be at the same resolution.
The final resolution of the H and K maps is 0.2$''$ and
that of the J map is 0.3$''$. From the flux density calibrated maps we
therefore calculated a J $-$ H color map at a resolution of 0.3$''$ and a H$-$K
color map at a resolution of 0.2$''$.
As a result over a 1.9$''$ diameter aperture centered on source $a$
the colors turned out to be fairly uniform with mean
colors corresponding to the values given in
Tab.~\ref{ww09} and color variations of $\le \pm$ 0.15.
The nucleus of source $a$ which can be clearly
distinguished in the individual images at the three wavelengths
does not appear particularly red or blue. This indicates that there
are little variations in extinction or in spectral type across the source.

{\bf PKS~0743-006:}
In Fig.~\ref{fig15}
 we show the colors of all sources in our field of view
towards the quasar PKS~0743-006 at $z=0.994$.
Compared to the cluster J1836.3CR this field is more sparsely
populated. Clearly not all of the objects have colors corresponding to
high-redshift galaxies. Two of the objects ($e, c$) are at a location in the
JHK color-color diagram that is populated by local giant or dwarf
stars. For comparison -- and as a convenient check of our calibration
-- we have also added the position of the bright reference star on
which the adaptive optics loop was locked. This star is of type A0,
in agreement with the measured colors.
All other objects are located close to the middle of the
evolutionary track for a passively evolving galaxy at a redshift of $z=0.994$.
This is consistent with an age closer to 2 Gyrs than 10 Gyrs. The age
may even be lower under the assumption of intrinsic reddening.
This would indicate that the light of these objects is dominated by
a reasonably young, blue stellar population.

Although the identification of these objects as $z=0.994$ galaxies has to
be confirmed spectroscopically,
an alternative explanation for the nature of these object is difficult
to find. Even taking the uncertainties of the measured colors into
account these sources are positioned well to the lower right of the
mean colors of spiral galaxies or those of local dwarfs and giant stars.
The only other objects that are located in this area of the
JHK-color diagram are local HII regions, if their near-infrared
emission is un-reddened and dominated by free-free radiation.
Having 3 of these objects in the same field as the quasar (with similar
colors) is very unlikely.

The best information on the sizes of the sources in this field
is from the data taken with a 0.05$''$/pixel sampling.
While the quasar itself is unresolved as compared to  radial and
tangential cuts with respect to the direction of the AO reference,
we find one of the brighter sources ($b$) that is close to the
quasar as clearly extended in all directions.
This source has an angular separation to the quasar of only 3.8$''$.
Deconvolving its measured size with the size of the unresolved quasar
by subtracting the values in quadrature its deconvolved source size
is $0.22'' \times 0.16''$ at a position angle of about 45$^\circ$.
At a redshift of $z=0.994$ this corresponds to a linear size of
1.9~kpc $\times$ 1.4~kpc.
The extent of the source, its small angular separation from the quasar,
and its colors make this object the best candidate for a cluster galaxy
which is associated with the quasar PKS~0743-006.
The previously described blue colors with respect to the values expected for
a passively evolving galaxy at a redshift of $z=0.994$ then imply that
we look at a galaxy with a region of enhanced star formation that
extends over an area of 2.7~kpc$^2$.

\subsection{Bulge sizes and structures in J1836.3CR}

\begin{figure*}
\resizebox{12cm}{!}{\includegraphics{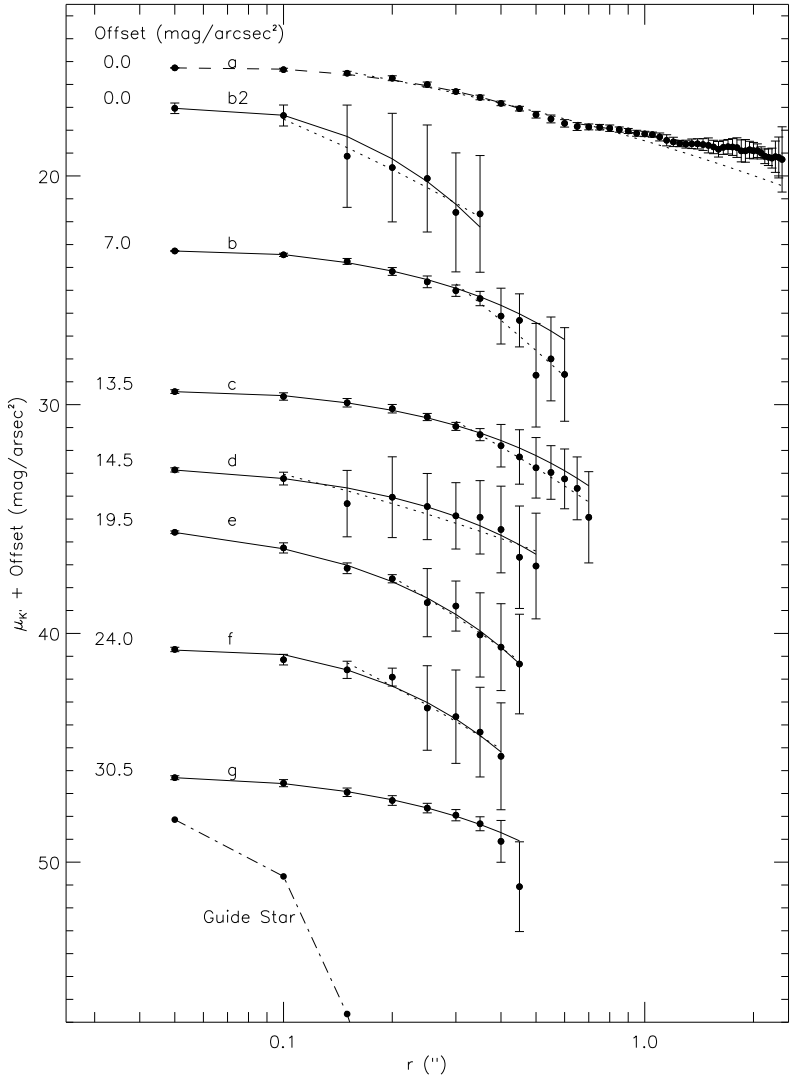}}
\hfill
\parbox[b]{55mm}{
\caption{Radial intensity profiles of the J1836.3CR cluster members.
performed on the deconvolved ADONIS/SHARP~II+ data taken in the K$'$
band with a 0.1$''$ pixel scale interpolated onto a 0.05$''$ pixel grid. The errors are
standard deviations of values at a given radius. The solid and dashed
lines are the result of a reduced $\chi^2$ fit of a model (see text)
to all data (solid line) and to data at larger radii only (dashed line,
De Vaucouleurs law). For comparison we have plotted the profile of the
AO guide star after the same number of iterations with the Lucy deconvolution algorithm (data connected
with a dash-dotted line).}
\label{fig16}}
\end{figure*}

Our best K-band AO images of J1836.3CR can  be used to derive
structural information on individual cluster members.
We determined radial profiles in the direct and deconvolved images.
The radial profiles of the deconvolved images have the advantage
that they have been corrected for the non-diffraction limited part of the
PSF and allow a much clearer view on the detailed distribution of
light in each object. In Fig.~\ref{fig16}
 we show the profiles of 8 sources
in the field including a profile of the guide star after the same
number of Lucy deconvolution iterations ($\sim 1000$).

\begin{figure}
\resizebox{\hsize}{!}{\includegraphics{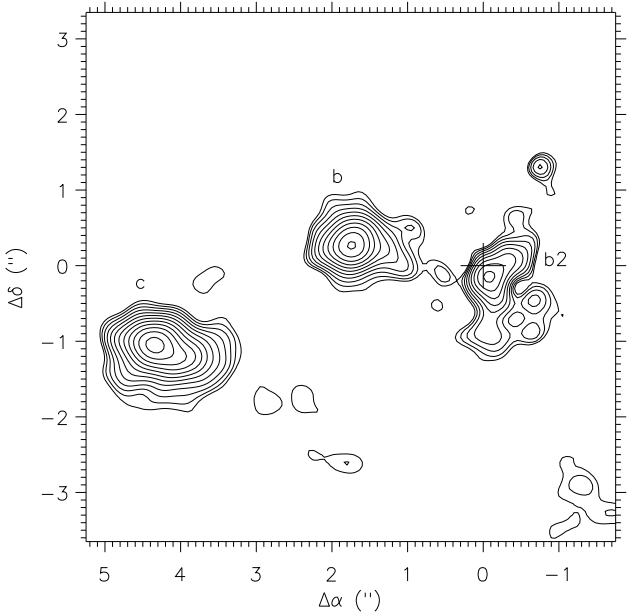}}
\caption{Subtraction of a radially averaged image of the central source $a$ in
J1836.3CR reveals the double nucleus structure of $a$.
The resultant K$'$-band image shown is based on 0.05$''$ pixel scale
AO-observations, smoothed to 0.3$''$ resolution after subtraction of $a$.
The lowest contour level
corresponds to 19.6~mag/arcsec$^2$; contour spacing is
0.3~mag/arcsec$^2$. The coordinate system is centered on the peak of
$a$ (cross).}
\label{fig17}
\end{figure}

To determine the profile of source $b$ we first subtracted a
radially averaged image of source $a$ as determined from its
non-contaminated section to the north. This process revealed a further,
even weaker source $b2$ (see Fig.~\ref{fig17}). Both sources $b$ and $b2$ were subtracted from the
image before calculating the average radial profile of $a$.

Following Kormendy (\cite{Kormendy}) the radial profiles were fitted
via a reduced $\chi^2$ test as
a combination of a $r^{-1/4}$ spheroid (De Vaucouleurs \cite{De Vaucouleurs})
with a surface brightness

\begin{equation}
\mu_{\rm V} = \mu_{\rm V}(0) + 8.325[(r/r_0)^{1/4} - 1]
\end{equation}
and an exponential intensity fall-off
with a sharp inner cutoff radius (Kormendy \cite{Kormendy}).
with a surface brightness

\begin{equation}
\mu_{\rm E} = \mu_{\rm E}(0) + 1.0857[\alpha r + (\beta/r)^3].
\end{equation}

Here $r_0$ and $\beta$ are the corresponding cutoff radii, 
$1/\alpha$ is the exponential scale length  and $\mu_{\rm V}(0)$ and $\mu_{\rm E}(0)$
are the central surface brightnesses for the $r^{-1/4}$ spheroid and the
exponential disk (if it had not been cut-off) in units of magnitudes per square arcseconds.
The surface brightness $\mu(r)$ and intensity distribution $I(r)$
are linked via $\mu (r) = -2.5 \times \log I(r)$.
None of the sources, with the exception of $a$, can be fit via an
$r^{-1/4}$ law alone.
The distributions at small radii are too flat
with respect to their decrease at larger radii;
they can only at larger radii
be approximated with a De Vaucouleurs law. This indicates
that these objects are not typical elliptical galaxies which are
usually found amongst the brightest cluster members
(Thuan and Romanishin \cite{Thuan}). All sources -- except object $a$ -- are best
fit with exponential profiles suggesting that they are more similar to
spiral disks. However, no central bulge could be detected with an upper size limit of
0.6--0.9~kpc.
This could be due to the fact that the angular resolution of 0.15$''$ is
still not sufficient to clearly discriminate between the bulge and
disk components.
Alternatively it could imply that the bulges are intrinsically  weak
with respect to the disks. The statistical investigation of bulge to disk
intensity ratios by Solanes et al. (\cite{Solanes}) would then suggest that
the galaxies in the center of J1836.3CR are of type Sa and S0 and that the
center of this cluster represents a dense environment.

Source $a$ shows a radial profile of a typical cD galaxy that is expected
to be present in the centers of rich galaxy clusters.
The cD galaxies in centers of clusters with low richness
can usually be described as giant ellipticals (Morgan et al. \cite{Morgan},
Albert et al. \cite{Albert}). However, cD galaxies in rich clusters
show excess flux density in their profiles at large radii with respect
to an $r^{-1/4}$ law (Oemler \cite{Oemler}, Dressler \cite{Dressler}).
This is also  the case for source $a$. The profile can only
be fit by a De Vaucouleurs law for radii $< 0.6''$. The resulting
scale length (see Tab.~\ref{ww11}) is in agreement with the lower limit of the
range of scale lengths of $r^{-1/4}$ laws found for the central
regions of bright cD galaxies in rich Abell clusters (Oemler
\cite{Oemler}).
The extended flux at larger radii of the central cD galaxy in J1836.3CR
may be indicative of either tidal disturbances due to galaxy-galaxy
interactions of cluster members in which the debris is sinking onto
the central massive galaxy (Gallagher \& Ostriker \cite{Gallagher}) or even
the result of recent merger processes (Ostriker \& Tremaine \cite{Ostriker}).
These activities are not likely to be important in clusters of
low richness (Thuan and Romanishin \cite{Thuan}) but can very well be responsible
for the structures  of objects in rich clusters.

\small
\begin{table}
\caption{\label{ww11} Disk parameters for the observed galaxies in
J1836.3CR. The mean error in the exponential scale length $1/\alpha$
is 0.40 kpc. The mean error in the inner cut-off radius $\beta$ is 0.07 kpc.
The mean error in the central disk surface brightness $\mu_c$ is
0.2~mag.} 
\begin{center}
\begin{tabular}{ccccccccccccc}
\hline\hline
source & $1/\alpha$ & $\beta$ & $\mu_c$ \\
&  (kpc) & (kpc) & (mag/arcsec$^2$) \\
\hline
a & 3.22 & 0.19 & 15.77 \\
b2 & 2.15 & 0.19 & 17.02 \\
b & 2.31 & 0.19 & 16.94 \\
c & 2.49 & 0.19  & 16.49 \\
d & 1.37 & 0.12 & 17.89 \\
e & 2.12 & -- & 17.35 \\
f & 2.30 & -- & 17.47 \\
g & 2.08 & 0.20 & 16.43 \\
\hline
\end{tabular}
\end{center}
\end{table}
\normalsize

\subsection{Luminosity of the cluster members}

In order to check how representative the individual galaxies that we
studied are we compared their luminosities to those of $L_*$ galaxies
at the corresponding redshifts of the three clusters at $z=0.0157$, $z=0.414$,
and $z=0.994$.
From a subsample of the optically selected Anglo-Australian
Redshift-Survey, Mobasher et al. (\cite{Mobasher})
find the best infrared Schechter luminosity function parameters as
$M_{\rm K}^{\;\;*} = -25.1 \pm 0.3$ with $\alpha = -1.0 \pm 0.3$ for
$H_0 = 50$~km~s$^{-1}$~Mpc$^{-1}$ and $q_0 = 0.02$.
From their sample they also find that E/S0 and spirals have identical
infrared luminosity functions within the errors.
Their parameters are very much in agreement with recent determinations
of $M_{\rm K}^{\;\;*}$ by De Propris et al. (\cite{De Propris}) and Gardner et al. (\cite{Gardner}).
For $h = H/H_0 = 0.5$ and $q_0$ = 0.5 they find
$M_{\rm K}^{\;\;*} = -24.8$ with bright galaxy slopes of $\alpha = -0.78$
and
$M_{\rm K}^{\;\;*} = -24.6 \pm$ 0.1 with $\alpha = -0.91 \pm$ 0.1, respectively.
For the purpose of comparison we use the same values
for $H_0$ and $q_0$ and obtain the following results for our targets:

{\bf Abell~262:}
For our low redshift cluster both objects we study here
-- UGC~1344 and UGC~1347 -- are $L_*$ galaxies to within less than 0.5
magnitudes.

{\bf J1836.3CR:}
For this cluster we find that the bright source $a$ is just
0.6 magnitude brighter and that the other sources are about 1 to
2 magnitudes fainter than a typical $L_*$ galaxy at that redshift of $z=0.414$.

{\bf PKS~0743-006:}
For the 3 presumed galaxies close to
the quasar PKS~0743-006 at $z=0.994$ we find that two are about
0.7 magnitudes brighter than an $L_*$ galaxy and the weakest one is 2
magnitudes fainter than $L_*$.
This is consistent with the statement that the brighter members
of the observed clusters are on the K-$z$ relation which, as emphasized
by Lilly (\cite{Lilly}), shows only a small scatter
of $\sigma \sim 0.3$~mag at redshifts below $z=1.5$.
In addition it should be mentioned that the quasar PKS~0743-006
itself shows very similar colors compared to the galaxy
candidates in the observed field. However, due to the
nuclear contribution it is -- as expected (see data in Dunlop et al. \cite{Dunlop}
and Lehnert et al. \cite{Lehnert}) -- between 1.5 and 2 mag
brighter than what is predicted by the galaxy K-$z$ relation.
We also find that the measured JHK fluxes of this variable quasar were
at the time we took the data about  0.7 to 1.0 magnitudes
below what is given in the literature (White et al. \cite{White}, L\'epine et
al. \cite{Lepine}).
Since quasar host galaxies have been found to be often as
bright  as $5 \times L_*$ (McLeod \& Rieke \cite{McLeod1995},
S\'anchez et al. \cite{Sanchez})
the underlying host galaxy may contribute substantially to the overall
flux density of the quasar in this low state.

This indicates that the source properties that we determined in the
previous sections are representative for typical $L_*$ cluster members
at the corresponding redshifts.

\section{Summary and Conclusions}

We have presented high angular resolution NIR observations
of three galaxy clusters at different redshifts using adaptive optics.
In the case of the barred spiral UGC~1347 in Abell~262 we presented
the first adaptive optics data using the laser guide star provided by
the ALFA system.

The diagnostic ratios for the nucleus of UCG~1347
indicate recent and ongoing star formation activity.
In addition to the
resolved NIR nucleus in UGC~1347 we found
a bright and compact region of recent and enhanced star formation at one tip of the bar.
The $L_{\rm K} / L_{\rm Lyc}$ ratio as well as the V~$-$~K color of that region
imply that a starburst happened
about 10$^7$ years ago.
For UGC~1344 we found that the overall star formation activity is low
and that the system is deficient in the fuel for star formation.

The comparison of seeing-corrected
nuclear bulge sizes of a sample of
26 cluster galaxies within and outside of the Abell radius of Abell~262
and Abell~1367 indicates that the
galaxies in the inner part of the cluster show a tendency for more compact
bulges than those outside.
This phenomenon can tentatively be ascribed to an increased star formation activity
due to interactions of cluster members inside the Abell radius.
Such an increase of central activity is also indicated at other
wavelengths.
Scodeggio and Gavazzi (\cite{Scodeggio}) find in a 21~cm survey of spiral galaxies
in clusters that about 30\% of them show extended
radio continuum emission and that a substantial fraction of those (but not all)
show indications of interaction.
Moss and Whittle (\cite{Moss1993}) find from an H$\alpha$ survey of cluster spirals
that interacting spirals show a strong tendency to have compact
nuclear H$\alpha$ emission which the authors conclude to be most likely
due to tidally induced star formation from galaxy-galaxy interactions,
since interactions are more likely to happen close to the cluster center.
Several successive high-speed encounters between galaxies
may lead to gas inflow and strong star
formation activity (Fujita \cite{Fujita}).
This would imply that dynamically induced star formation is more important
in the center than the outer parts of a cluster, although current
investigations have not yet convincingly shown the obvious presence of such
an correlation.
Future observations of larger samples of cluster members are clearly needed
to substantiate these correlations.

Since the spiral content of galaxy clusters at higher redshifts
is about as large as  the spiral content in the field at $z=0$
(Oemler et al. \cite{Oemler}),
detailed observations of galaxies in low-redshift spiral-rich clusters
may provide essential information of the
cluster evolution at higher redshifts in general. In particular it
would allow us to study in detail the influence of galaxy harassment.
Abell~262 and  Abell~1367 are spiral-rich clusters. With a ratio of
spirals to S0 and E0 galaxies of 47\%/53\% and 43\%/57\%, respectively,
their spiral content is similar to that in the field at $z=0$ of 55\%/45\%
and to that in clusters at a redshift of $z=0.4$ of 40\%/60\%
~(Oemler \cite{Oemler}).

UGC~1347 and UGC~1344 could very well be taken as examples of the
blue and red fraction of clusters at higher redshift.
Couch et al. (\cite{Couch}) reports on 3 galaxy clusters at $z=0.3$ measured with
the HST.
In these clusters he finds the fraction of spirals at least 3
times higher than at $z=0$ which would approximately correspond to
the spiral fraction of Abell~262 and Abell~1367.
About 20\% of all galaxies show signs of interaction.
The blue fraction of the cluster population
shows morphologies similar to Sb-Sdm/Irr galaxies
with compact knotty regions of star formation.
These knots may be very similar to the bright star formation region we
found for UGC~1347.
The red fraction of the cluster population they find
is 1-2 Gyr past the last major star formation event
and has morphological similarities to S0-Sb disks.
This may be very similar to UGC~1344 and
source $a$ in J1836.3CR.

However, the radial profile of source $a$ in J1836.3CR shows
indications for recent or ongoing interactions between
cluster members. Here the enhanced flux above the De Vaucouleurs
fit to the data at larger radii identifies source $a$ as the central
cD galaxy in a rich cluster environment in which interactions
between cluster members are probably still of importance for their
further evolution.

From an investigation of NIR colors
Hutchings \& Neff (\cite{Hutchings}) find for 5 quasars at redshifts ranging from
$z=0.06$ to 0.3 that they are located in mostly evolved groups of
galaxies with an indication for average extinction values of $A_{\rm V}$ = 2--3~mag
quite similar to J1836.3CR at $z=0.414$ discussed in this paper.
However, on the basis of large-scale investigations (Butcher \&
Oemler, \cite{Butcher}, Ellis \cite{Ellis})
as well as studies of individual galaxy clusters
(e.g. Morris et al. \cite{Morris}) one finds that
both in clusters and in the field the blue fraction of
galaxies generally increases towards higher redshifts
(the Butcher Oemler effect, BOE).
This is consistent with our finding of relatively blue NIR
colors in the galaxy candidates associated with
PKS~0743-006 at $z=0.994$.

Based on an investigation of Abell~2390 at $z=0.2279$
and a comparison to the cluster MS~1621.5+2640 at $z=0.4274$
Abraham et al. (\cite{Abraham}) and Morris et al. (\cite{Morris})
suggest that the star formation process is shut down by a
combination of gas stripping followed by gas exhaustion via star formation.
This truncation of star formation activity may
explain both the BOE and the large fraction of S0 galaxies in clusters.
This also suggests that  truncated star formation induced by infall
does not play a major role in driving cluster galaxy
evolution at lower redshifts although this mechanism may have played a role
in earlier history.
The barred structure of UGC~1347 as well
as the strong recent star formation event at one tip of
the bar may indicate that this object is well suited to study the
corresponding physical processes of cluster galaxy evolution in great detail.

\begin{acknowledgements}
We are grateful to the Calar Alto and La Silla staff for their excellent support
and hospitality.
We thank M. Lehnert, H.W. Rix, R. Genzel, L. Tacconi and T. Boller
for helpful discussions.
We are particularly grateful to P. Amram  and M. Marcelin who kindly provided
the H$\alpha$ continuum and line images to us. We furthermore thank the 
referee for very helpful and constructive suggestions.
\end{acknowledgements}


\begin{thebibliography}{}
\bibitem[1996]{Abraham} Abraham, R. G., Smecker-Hane, T.A., Hutchings,
J.B., et al., 1996, \apjj{471}{694}
\bibitem[1977]{Albert} Albert, C., White, R., Morgan, W.W., 1977, \apjj{211}{309}
\bibitem[1994]{Amram} Amram, P., Marcelin, M., Balkowski, C., Cayatte, V.,
et al., 1994, \asas{103}{5}
\bibitem[1994]{Arsenault} Arsenault, R., Salmon, D.A., Kerr, J. M., et
  al., 1994. In: Adaptive Optics in Astronomy, Ealey, M.A., Merkle,
  F. (eds.), Proc. SPIE 2201, 833
\bibitem[1994]{Beuzit} Beuzit, J.-L., Hubin, N., Gendron, E., et al.,
1994. In: Adaptive Optics in Astronomy, Ealey, M.A., Merkle, F. (eds.), Proc. SPIE 2201, 955
\bibitem[1997]{Bravo-Alfaro} Bravo-Alfaro, H., Szomoru, A.,
Cayatte, V., et al., 1997, \asas{126}{537}
\bibitem[1993]{Bruzual} Bruzual, A.G., Charlot, S., 1993, \apjj{405}{538}
\bibitem[1984]{Butcher} Butcher, H., Oemler, A., 1984, \apjj{285}{426}
\bibitem[1996]{Condon1996} Condon, J.J., Cotton, W.D., Greisen, E.W., et al., 1996, NCSA Astronomy Digital Image Library
\bibitem[1992]{Condon1992} Condon, J.J., 1992, \anrev{30}{575}
\bibitem[1998]{Couch} Couch, W.J., Berger, A.J., Smail, I., et al., 1998, \apjj{497}{188}
\bibitem[1999]{Davies1999} Davies, R.I., Hackenberg, W., Ott, T., et al., 1999, \asas{138}{345}
\bibitem[1998]{Davies1998} Davies, R.I., Hackenberg, W.,  Ott, T., et al.,
1998. In: Adaptive Optics System Technologies, Bonaccini, D., Tyson,
R.K. (eds.), Proc. SPIE 3353, 116
\bibitem[1994]{Doyon} Doyon, R., Joseph, R.D., Wright, G.S., 1994, \apjj{421}{101}
\bibitem[1979]{Dressler} Dressler, A., 1979, \apjj{231}{659}
\bibitem[1993]{Dunlop} Dunlop, J.S., Taylor, G.L., Hughes, D.H., Robson, E.I., 1993
   \mn{264}{455}
\bibitem[1996]{Eckart1996} Eckart , A., Cameron, M., Boller, Th., Krabbe, et al., 1996, \apjj{472}{588-599}
\bibitem[1990]{Eckart1990} Eckart, A., Cameron, M., Rothermel, et al., 1990, \apjj{363}{451}
\bibitem[1998]{Eisenhauer} Eisenhauer F., Quirrenbach A., Zinnecker, H., Genzel, R., 1998, \apjj{498}{278}
\bibitem[1997]{Ellis} Ellis, R., 1997, \anrev{35}{389}
\bibitem[1998]{De Propris} De Propris, R., Eisenhardt, P.R., Stanford, S.A.,
   Dickinson, M., 1998, \apjj{503}{L48}
\bibitem[1959]{De Vaucouleurs} de Vaucouleurs, G., 1959, Handbuch der
Physik 53, 311
\bibitem[1987]{De Vries} De Vries, H.W., Heithausen, A., Thaddeus, P., 1987, \apjj{319}{723}
\bibitem[1989]{Draine} Draine, B.T., 1989. In: Infrared Spectroscopy in
    Astronomy, Kaldeich, B.H. (ed.), Proc. of ESA, ESA-SP 290
\bibitem[1998]{Drummond} Drummond, J.D., Fugate, R.Q.,  Christou,
J.C., Hege, K., 1998, Icarus 132, 80
\bibitem[1985]{Fairclough} Fairclough, J.H., 1985. In: Extragalactic
Infrared Astronomy, Gondhalekhar, P.M. (ed.), Rutherford Appelton
Laboratory Workshop, RAL-85-086 
\bibitem[1981]{Fanti} Fanti, C., Mantovani, F., Tomasi, P., 1981, \asas{43}{1}
\bibitem[1978]{Frogel} Frogel, J.A., Persson, S.E., Aaronson, M., Matthews, K.,
    1978,\apjj{220}{75}
\bibitem[1998]{Fujita} Fujita, Y., 1998, \apjj{509}{587}
\bibitem[1997]{Gardner} Gardner, J.P., Sharples, R.M., Frenk, C.S., Carrasco, B.E.,
   1997, \apjjl{480}{99}
\bibitem[1972]{Gallagher} Gallagher, J.S., Ostriker, J.P., 1972, \ajj{77}{288}
\bibitem[1996]{Gavazzi1996a} Gavazzi, G., Boselli, A., 1996, Astrophys. Letters \& Communcations 35, 1
\bibitem[1996]{Gavazzi1996b} Gavazzi, G., Piierini, D., Baffa, et al., 1996, \asas{120}{521}
\bibitem[1990]{Gavazzi1990} Gavazzi, G., Trinchieri, G., Boselli, A., 1990, \asas{86}{109}
\bibitem[1983]{Gehrz} Gehrz, R.D., Sramek, R.A., Weedman, D.W., 1983, \apjj{267}{551}
\bibitem[1995]{Genzel} Genzel, R., Weitzel, L., Tacconi-Garman, et al., 1995, \apjj{444}{129}
\bibitem[1985]{Giovanelli1985} Giovanelli, R., Haynes, M.P., 1985, \ajj{90}{2445}
\bibitem[1982]{Giovanelli1982} Giovanelli, R., Haynes, M.P., Chincarini, G.L., 1982, \apjj{262}{442}
\bibitem[1997]{Glindemann} Glindemann A., Hamilton, D., Hippler, S., et
  al., 1997. In: Proc. of the ESO Workshop on Laser Technology for Laser Guide Star
  Adaptive Optics Astronomy, N. Hubin (ed.), 120
\bibitem[1981]{Gregory} Gregory, P.C., Thompson, L.A., Tifft, W.G., 1981, \apjj{243}{411}
\bibitem[1978]{Heckman} Heckman, T.M., Balick, B. Sullivan III, W.T. 1978, \apjj{224}{745}
\bibitem[1986]{Helou} Helou, G., 1986, \apjjl{311}{33}
\bibitem[1993]{Hewitt} Hewitt, A., Burbidge, G., 1993, \apjs{87}{451}
\bibitem[1983]{Hildebrand} Hildebrand, R.H., 1983,
Roy. Astron. Soc. Quart. J. 24, 267
\bibitem[1998]{Hippler} Hippler, S., Glindemann, A., Kasper, M., Kalas,
P., et al., 1998. In: Adaptive Optical System Technologies,
Bonaccini, D., Tyson, R.D. (eds.), Proc. SPIE 3353, 44 
\bibitem[1992]{Hofmann} Hofmann, R., Blietz, M., Duhoux, P., et al.,
    1992. In: Progress in Telescope and Instrumentation Technologies,
    Ulrich, M.H. (ed.), ESO Conference and Workshop Proceedings No.42,
    617
\bibitem[1997]{Hubin} Hubin, N., 1997. In: Optical Telescopes of
      Today and Tomorrow, Ardeberg, A.L. (ed.), Proc. SPIE 2871, 827
\bibitem[1997]{Hutchings} Hutchings, J.B., Neff, S.G. , 1997, \apjj{113}{550}
\bibitem[1991]{Jackson} Jackson, J.M., Eckart, A., Cameron, M., et al., 1991, \apjj{375}{105}
\bibitem[1994]{Kennicutt} Kennicutt, R.C., Tamblyn, P., Congdon, C.E., 1994, \apjj{435}{22}
\bibitem[1977]{Kormendy} Kormendy, J., 1977, \apjj{217}{406} 
\bibitem[1994]{Krabbe} Krabbe, A., Sternberg, A., Genzel, R., 1994, \apjj{425}{72}
\bibitem[1978]{Larson} Larson, R. B., Tinsley, B. M., 1978, \apjj{219}{46}
\bibitem[1996]{Leitherer} Leitherer, C., 1996. In: ASP conf. series vol. 98, 373
\bibitem[1992]{Lehnert} Lehnert, M.D., Heckmann, T.M., Chambers, K.C., Miley, G.K.,
   1992, \apjj{393}{68}
\bibitem[1985]{Lepine} L{\'e}pine, J.R.D., Braz, M.A., Epchtein, N., 1985, \asa{149}{351}
\bibitem[1988]{Lilly} Lilly, S.J., 1988, \apjj{340}{77}
\bibitem[1985]{Lonsdale} Lonsdale, C.H., Helou, G., Good, J.C., Rice, W., 1985,
   Cataloged Galaxies and Quasars Observed in the IRAS Survey,
   NASA-JPL, Pasadena
\bibitem[1974]{Lucy} Lucy, L.B., 1974, \ajj{79}{745}
\bibitem[1991]{Mas-Hesse} Mas-Hesse, J.M., Kunth, D., 1991, \asas{88}{399}
\bibitem[1997]{Max} Max, C.E., Olivier, S.S., Friedman, et al., 1997, Sci 277, 1649
\bibitem[1995]{McLeod1995} McLeod, K.K., Rieke, G.H., 1995, \apjjl{454}{77}
\bibitem[1977]{Melnick} Melnick, J., Sargent, W.L.W., 1977, \apjj{215}{401}
\bibitem[1994]{Meynet} Meynet, G., Maeder, A., Schaller, et al., 1994, \asas{103}{97}
\bibitem[1993]{Mobasher} Mobasher, B., Sharples, R.M., Ellis, R.S., 1993, \mn{263}{560}
\bibitem[1996]{Moore} Moore, B., Katz, N., Lake, G., et al., 1996, Nat 379, 613
\bibitem[1975]{Morgan} Morgan, W.W., Kayser, S., White, R., 1975, \apjj{199}{545}
\bibitem[1998]{Morris} Morris, S.L., Hutchings, J.B., Carlberg, R.G., et
al., 1998, \apjj{507}{84}
\bibitem[1993]{Moss1993} Moss, C., Whittle, M., 1993, \apjjl{407}{17}
\bibitem[1977]{Moss1977} Moss, C., Dickens, R.J., 1977, MNRAS 178,701
\bibitem[1997]{Oemler} Oemler, Dressler, Butcher, 1997, \apjj{474}{561}
\bibitem[1993]{Oly} Oly, C., Israel, F.P., 1993, \asas{100}{263}
\bibitem[1989]{Osterbrock} Osterbrock, D.E., 1989, Astrophysics of Gaseous
Nebulae and Active Nuclei, University Science Books, Mill Valley
\bibitem[1975]{Ostriker} Ostriker, J.P., Tremaine, S.D., 1975, \apjjl{202}{113}
\bibitem[1997]{Quirrenbach} Quirrenbach, A., Hackenberg, W., Holstenberg,
H.C., Wiln\-hammer, N., 1997. In: Adaptive Optics and Applications,
Tyson, R.K., Fugate, R. (eds.), Proc. SPIE 3126, 35 
\bibitem[1993]{Rieke1993} Rieke, G. H., Loken, K., Rieke, M. J., Tamblyn, P., 1993, \apjj{412}{99}
\bibitem[1980]{Rieke1980} Rieke, G. H., Lebofsky, M. J., Thompson, R. I., et al., 1980, \apjj{238}{24}
\bibitem[1955]{Salpeter} Salpeter, E. E., 1955, \apjj{121}{161}
\bibitem[1997]{Sanchez} S{\'a}nchez, S.F., Conz\'alez-Serrano, J.I.,
   Carballo, et al., 1997. In: Proceedings of the ESO/IAC Workshop on
   Quasar Hosts, Clements, D.L., P\'erez-Fournon, I. (eds.), 21
\bibitem[1993]{Schaerer} Schaerer, D., Maynet, G., Maeder, A., Schaller, G., 1993, \asa{274}{1012}
\bibitem[1997]{Schinnerer} Schinnerer, E., Eckart, A., Quirrenbach, et al., 1997, \apjj{488}{174}
\bibitem[1993]{Scodeggio} Scodeggio, M., Gavazzi, G., 1993, \apjj{409}{110}
\bibitem[1978]{Shostak} Shostak, G.S. 1978, \asa{68}{321}
\bibitem[1989]{Solanes} Solanes J.M., Salvador-Sole, E., Sanroma, M., 1989, \ajj{98}{798}
\bibitem[1997]{Stanghellini} Stanghellini, C., O'Dea, C.P., Baum, et al., 1997, \apjj{325}{943}
\bibitem[1996]{Tacconi-Garman} Tacconi-Garman, L.E., Sternberg, A., Eckart, A., 1996, \ajj{112}{918}
\bibitem[1994]{Tammann} Tammann, G.A., Loeffler, W., Schroeder, A., 1994, \apjs{92}{487}
\bibitem[1981]{Thuan} Thuan, T.X., Romanishin, W., 1981, \apjj{248}{439}
\bibitem[1993]{Tornikoski} Tornikoski, M., Valtaoja, E., Terasranta, H.,
et al., 1993, \ajj{105}{1680}
\bibitem[1992]{Wainscoat} Wainscoat, R.J., Cowie, L.L., 1992, \ajj{103}{332}
\bibitem[1980]{Wilkerson} Wilkerson, M.S., 1980, \apjj{240}{115}
\bibitem[1988]{White} White, G.L., Jauncey, D.L., Savage, A., et al., 1988, \apjj{327}{561}
\bibitem[1981]{Whittet} Whittet, D.C., 1981, Quarterly J. Roy. Astron. Soc. 22, 3
\bibitem[1988]{Wunderlich} Wunderlich, E., Klein, U., 1988, \asa{206}{47}
\bibitem[1991]{Young} Young, J.S., Scoville, N., 1991, \anrev{29}{581}
\end{thebibliography}
\end{document}